\documentclass[runningheads,fleqn]{svmult}

\usepackage{makeidx}   
\usepackage{graphicx}  
\usepackage{subeqnar}  
\usepackage{multicol}  
\usepackage{physmult}  
\makeindex             

\usepackage{latexsym}
\usepackage{amssymb}
\usepackage{amsbsy}
\usepackage[mathscr]{eucal}
\newcommand \beq{\begin{equation}}
\newcommand \eeq{\end{equation}}

\newcommand \bea{\begin{eqnarray}}
\newcommand \eea{\end{eqnarray}}

\def\0{\over } \def\2{{1\over2}} \def\4{{1\over4}}
\def\5{\hat } \def\6{\partial }

 \def\b{\beta }  \def\d{\delta }

 \def\o{\omega }
   
\def\O{\Omega }

\def\({\left(} \def\){\right)} \def\<{\langle } \def\>{\rangle }

\newcommand{\nn}{\nonumber\\ }
\def\bp{\begin{picture}(0,0)}
\def\ep{\end{picture}}

\def\Tr{{\,\mathrm Tr\,}}
\def\Im{\rm Im}
\def\Re{\rm Re}

\def\del{\partial }

\def\simge{\mathrel{%
    \rlap{\raise 0.511ex \hbox{$>$}}{\lower 0.511ex \hbox{$\sim$}}}}
\def\simle{\mathrel{
    \rlap{\raise 0.511ex \hbox{$<$}}{\lower 0.511ex \hbox{$\sim$}}}}

\def\grad{\nabla}                               
\def\del{\partial}                              

\def\frac#1#2{{#1 \over #2}}

\def\half{\ifinner {\scriptstyle {1 \over 2}}
    \else {1 \over 2} \fi}

\def\bra#1{\langle#1\vert}              
\def\ket#1{\vert#1\rangle}              

\def\simge{\mathrel{%
    \rlap{\raise 0.511ex \hbox{$>$}}{\lower 0.511ex \hbox{$\sim$}}}}
\def\simle{\mathrel{
    \rlap{\raise 0.511ex \hbox{$<$}}{\lower 0.511ex \hbox{$\sim$}}}}

\def\slashchar#1{\setbox0=\hbox{$#1$}           
    \dimen0=\wd0                                 
    \setbox1=\hbox{/} \dimen1=\wd1               
    \ifdim\dimen0>\dimen1                        
       \rlap{\hbox to \dimen0{\hfil/\hfil}}      
       #1                                        
    \else                                        
       \rlap{\hbox to \dimen1{\hfil$#1$\hfil}}   
       /                                         
    \fi}                                         %

\def\subrightarrow#1{
   \setbox0=\hbox{
     $\displaystyle\mathop{}
     \limits_{#1}$}
   \dimen0=\wd0
   \advance \dimen0 by .5em
   \mathrel{
     \mathop{\hbox to \dimen0{\rightarrowfill}}
        \limits_{#1}}}                           

\def\nbfgrad{\mbox{\boldmath$\grad$}}

\def\bfnabla{\mbox{\boldmath$\nabla$}}
\def\bfsigma{\mbox{\boldmath$\sigma$}}

\def\tilQ{\mbox{v\cdot q}}
\def\tilQ1{\mbox{$v\cdot q_1$}}
\def\tilQ2{\mbox{$v\cdot q_2$}}

\newenvironment{petitchar}{\begin{list}{}
{\leftmargin1.5em\rightmargin0.0cm}%
\item\small}{\end{list}}

\def\journal#1#2#3#4{\ {#1}{\bf #2} ({#3})\  {#4}}

\def\AnnPhys{\journal{Ann.\ Phys.}}

\def\NPB{\journal{Nucl.\ Phys.\ {\bf B}}}

\def\PLB{\journal{Phys.\ Lett.\ {\bf B}}}

\def\PRC{\journal{Phys.\ Rev.\ {\bf C}}}
\def\PRD{\journal{Phys.\ Rev.\ {\bf D}}}
\def\PRL{\journal{Phys.\ Rev.\ Lett.}}
\def\PhysRept{\journal{Phys.\ Repts.}}

\def\SovPhysJETP{\journal{Sov.\ Phys.\ JETP}}

\def\SovJNuclPhys{\journal{Sov.\ J.\ Nucl.\ Phys.}}

\begin{document}
\title*{Theory of the Quark-Gluon Plasma}
\toctitle{Theory of the Quark-Gluon Plasma}
%
%
\titlerunning{Theory of the quark-gluon plasma}
%
\author{Jean-Paul Blaizot}
\institute{
Service de Physique Th\'eorique, CEA Saclay\\
91191 Gif-sur-Yvette Cedex, France
}

\authorrunning{Jean-Paul Blaizot}
%
%

\maketitle              

\bp\put(243,150){\large\tt SPhT-T01/074}\ep

\section{Introduction}

In spite of what the title might suggest, I shall not try to cover in these
lectures all interesting aspects of the theory of the quark-gluon
plasma. I shall rather focus on 
 progress made in recent years in understanding the high
temperature phase of QCD by using 
weak coupling techniques. Such techniques go far beyond strict perturbation
theory viewed as an  expansion  in powers of the gauge coupling. In fact
such an expansion becomes meaningless as soon as the coupling is not
vanishingly small. However, we shall see that a rather simple structure
emerges from weak coupling studies, with a characteristic hierarchy of scales
and degrees of freedom. The interactions renormalize the properties of these
elementary degrees of freedom, but does not destroy the 
simple picture of the high temperature quark-gluon plasma as a system of weakly
interacting quasiparticles. As we shall see at the end of these lectures, this
picture is supported by a first principle calculation of  the entropy which
reproduces accurately lattice data above 2 or 3 times the critical
temperature. 
 
Some of the material presented here is  borrowed from the recent
review~\cite{Blaizot:2001nr}, and  complements can also  be found in
\cite{Blaizot1997,Blaizot:1996ns,BIO95,Puri94,Coree91}. Another perspective
on some of the topics discussed here can be found in the lectures by A.~Rebhan.

The outline of the lectures is the following. In order to get a
first rough picture of the phase diagram of hadronic matter I use the bag
model to describe the quark-hadron phase transition: this exercise will
give us some familiarity with the thermodynamics of massless, non-interacting,
particles. Then I briefly recall  some  techniques of quantum field theory at
finite temperature needed to treat the interactions
\cite{KB62,Abrikosov63,Fetter71,BR86,Kapusta,LeBellac96}, and
introduce the concept of effective theory in a simple case of a
scalar field. Then I proceed to an analysis of the various important
scales and degrees of freedom of the quark-gluon plasma and focus on
the effective theory for the collective modes which develop at the
particular momentum scale $gT$, where $g$ is the gauge coupling and
$T$ the temperature. A powerful technique to construct the
effective theory is based on kinetic equations which govern the
dynamics of the hard degrees of freedom. Some of the collective
phenomena that are described by this effective theory are briefly
mentioned. Then I turn to the calculation of the entropy and show how
the information coded in the effective theory can be exploited in
(approximately) self-consistent
calculations \cite{Blaizot:1999ip,Blaizot:1999ap,Blaizot:2001fc}.

\section{The quark-hadron transition in the bag model.}

\begin{figure}
\includegraphics[width=.5\textwidth]{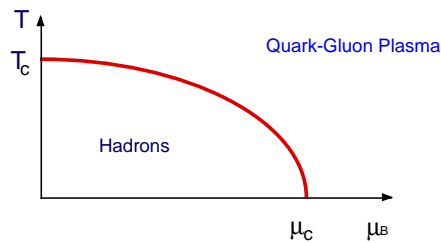}
\caption{The expected phase diagram of hot and dense hadronic matter 
        in the plane $\left(\mu_ B,T \right) $, where $T $ 
        is the temperature and $ \mu_ B $ the baryon  chemical potential}
\label{fig:phase_diag}
\end{figure}

The phase diagram of dense hadronic matter has the expected shape indicated
in Fig.~\ref{fig:phase_diag}.  There is a low density, low
temperature region, corresponding to the world of ordinary hadrons,
and a high  density, high temperature region, where the dominant
degrees of freedom are  quarks and gluons. The
precise determination of the transition line  requires
elaborate non perturbative techniques, such as those of lattice gauge
theories (see the lectures by F.~Karsch).  But one can get rough orders of
magnitude for the transition temperature and  density using a simple 
model dealing
mostly with non-interacting particles
\cite{Blaizot:1996ns,Puri94}.

Let us first consider  the transition  in the
  case  where $ \mu_B =0$.
At low temperature  this baryon free matter is composed of the lightest mesons,
i.e. mostly the pions. At sufficiently  high temperature one should also
take into account heavier mesons, but in  the present discussion
this is an inessential complication. We shall even  make a further 
approximation by treating the pion as a
massless particle. At very  high temperature, we shall consider that 
hadronic matter is composed
only of  quarks and antiquarks (in equal numbers), and gluons,
forming a quark-gluon  plasma\index{quark-gluon  plasma}. In both the high
temperature and the low temperature phases,  interactions are neglected
(except for the  bag constant to be introduced below).
The description of the transition will therefore  be dominated by
entropy\index{entropy} considerations, i.e. by counting  the degrees of 
freedom.

The energy density
$ \varepsilon $ and the pressure $ P $ of a
gas of massless pions are given by:
\beq
  \varepsilon  = 3\cdot  {\pi^ 2 \over 30} T^4\, ,\ \ \ \ \ \ P
= 3\cdot{ \pi^ 2 \over 90} T^4, \eeq where the factors 3 account
for the 3 types of pions $ (\pi^ +, $ $ \pi^ -, $
$ \pi^ 0). $

The energy density and pressure of the quark-gluon plasma are given
by similar  formulae:
\bea
  \varepsilon &  =& 37\cdot{ \pi^ 2 \over 30} T^4+B, \nonumber\\ P & =
& 37\cdot{ \pi^ 2 \over 90} T^4 - B,
  \eea where $ 37 = 2\times 8+{7 \over 8}\times 2\times 2\times
2\times 3 $ is the effective number of degrees of freedom\index{degrees of
freedom}  of gluons (8 colors, 2 spin states) and quarks (3 colors, 2 spins, 2
flavors, $ q $ and $ \bar q) $. The quantity $ B $, which is  added to the
energy density, and subtracted from the pressure, summarizes
interaction effects which are  responsible for a change in the
vacuum structure between the low temperature  and the high
temperature phases. It was introduced first in the \lq\lq bag
model\rq\rq\  of hadron structure as a restoring force needed to
equilibrate the pressure  generated by the kinetic energy of the
quarks inside the bag \cite{bagmodel}\index{bag model}. Roughly, the  energy of
the bag is
\beq
  E(R) = {4\pi \over 3} R^3B + {C \over R},
\eeq
where $ C/R$ is the kinetic energy of massless quarks.
Minimizing with respect to $ R, $ one finds that the energy at
equilibrium is
$ E \left(R_0 \right)=4B V_0, $  where $ V_0 = 4\pi R_0^3/3 $ is
the equilibrium volume. For a proton with $ E_0
\approx  1$ GeV and
$ R_0\approx 0.7$ fm, one finds $ E_0/V_0\simeq 0.7$
GeV/fm$^3 $, which corresponds to a \lq\lq bag constant\rq\rq\ $ B
\approx  175$ MeV/fm$^3$, or $ B^{1/4} \approx  192$ MeV.

\begin{figure}
\label{fig:pression}
\includegraphics[width=.5\textwidth]{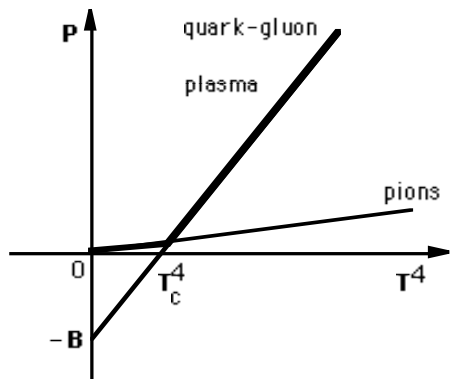}
\caption{The pressure of the massless pion gas compared to that of a 
quark-gluon
plasma, showing the transition temperature $T_c$.}
\end{figure}

We can now compare the two phases as a function of the temperature.
Fig.~\ref{fig:pression} shows how $ P $ varies as a
function of $ T^4. $ One sees that there exists  a transition
temperature
\beq\label{Tc}
  T_c= \left({45 \over 17\pi^ 2} \right)^{1/4}\ B^{1/4} \approx
0.72\ B^{1/4}, \eeq beyond which the quark-gluon plasma is
thermodynamically favored (has largest  pressure) compared to the
pion gas. For $ B^{1/4}\approx 200$ MeV, $ T_c \approx
150$ MeV.

\begin{figure}
\includegraphics[width=.5\textwidth]{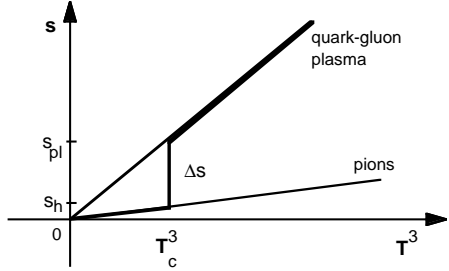}
\caption{The entropy density. The jump $ \Delta s $ at the transition is
proportional to the increase in the number of active degrees of freedom }
\label{fig:entropden}
\end{figure}

The variation of the entropy
density\index{entropy density} $ s =
\partial P /\partial T $  as a function of the temperature  is
displayed in Fig.~\ref{fig:entropden}. Note that  the bag constant $ 
B $ does not
enter explicitly the expression of the entropy. However, $ B $
is involved in  Fig.~\ref{fig:entropden} indirectly, via the
temperature $ T_c $ where the  discontinuity $ \Delta s $ occurs.
One verifies easily that the jump in entropy density
$ \Delta s = \Delta \varepsilon /T_c $ is directly proportional to
the change in the number of active  degrees of freedom when $ T $
crosses $ T_c $.

In order to extend these considerations to the case where
$\mu_B\ne 0$, we note that the transition is taking place when
the total pressure approximately vanishes, that is when the kinetic
pressure of quarks and gluons approximately equilibrates the bag
pressure. Taking as a criterion for the phase transition the
condition $P=0$, one replaces the value (\ref{Tc}) for $T_c$ by the
value $(90/37\pi^2)^{1/4}B^{1/4}\approx 0.70 B^{1/4}$, which is
nearly identical to (\ref{Tc}). We shall then assume that for any
value of $\mu_B$ and $T$, the phase transition occurs when
$P(\mu_B,T)=B$, where $B$ is the bag constant and $P(\mu_B,T)$
is the kinetic pressure of quarks and gluons:
\beq
P(\mu_B,T)=\frac{37}{90}\pi^2T^4+\frac{\mu_B^2}{9}(T^2+\frac{\mu_B^2}{9\pi^2}).
\eeq
The transition line is then given by $P(\mu_c,T_c)=B$, and it has
indeed the shape illustrated in Fig.~\ref{fig:phase_diag}.

The model that we have just described reproduces some of the bulk
features of  the equation of state obtained through lattice gauge
calculations (see the lecture by F.~Karsch). In particular, it exhibits the
characteristic increase of the entropy  density\index{entropy density} at the
transition which corresponds to the emergence of a large  number of new
degrees of freedom associated with quarks and gluons. Its
simplicity has made it popular for instance among the practitioners of
hydrodynamic  calculations with which one tries to simulate the
 behavior of matter  produced in high energy nuclear
collisions. As such it has been very useful.  One should be
cautious however when attempting to draw too detailed conclusions
about the nature of the phase transitions from such simple models. In 
particular
  this model  predicts
(by construction!) a discontinuous
transition; but this prediction should not be trusted. Further discussion of
this model can be found in \cite{Blaizot:1996ns}

\section{Quantum Fields at Finite Temperature }

The effects of interactions among quarks and gluons at finite temperature can be
calculated by using the tools of quantum field theory at finite temperature. We
shall briefly recall some essential formalism, and emphasize in particular the
periodicity properties of the propagators. At the end of this section we
discuss,  with a simple example of a scalar field, the method of effective field
theory which proves useful in problems where various scales can be separated.
In the example that we shall consider, the separation of scale is provided by
the Matsubara frequencies. As we shall see, in some cases, one is lead to
single out the mode with vanishing Matsubara frequency. The corresponding
effective theory is a classical field in three dimensions, and  the procedure
commonly called \lq dimensional reduction'\index{dimensional reduction}.

\subsection{Finite Temperature calculations}

All thermodynamic observables can be deduced from the partition
function\index{partition function}:
\beq
{\cal Z}={\rm tr\;}e^{-\beta H}.
\eeq
Thus the energy density\index{energy density} and the pressure\index{pressure}
are given by:
\beq
\epsilon=-\frac{1}{V}\frac{\del}{\del\beta}\ln Z\qquad\qquad
P=\frac{1}{\beta}\frac{\del}{\del V}\ln Z.
\eeq
In order to calculate the partition function, one may  observe that $e^{-\beta
H}$ is like an evolution operator in imaginary time\index{imaginary time}:
\beq
t\rightarrow -i\beta\qquad\qquad e^{-iHt}\rightarrow e^{-\beta H} .
\eeq
One may then take advantage of all the techniques developed to  evaluate
matrix elements of the evolution operator in quantum mechanics or field theory.

For instance one may use a perturbative
expansion\index{perturbative expansion}. We assume that one can split the
hamiltonian into
$ H=H_0+H_1 $
with $H_1\ll H_0$, and  define the following
``interaction
representation''\index{interaction representation} of the perturbation $H_1$:
\beq
  H_1(\tau)=e^{\tau H_0}H_1 e^{-\tau
H_0},
\eeq
  and similarly for other operators.
Using standard techniques, one can then obtain the following expression for the
partition function ${\cal Z}$\index{partition function}:
\beq\label{Z_pert}
{\cal Z}={\cal Z}_0\,\langle {\rm T}\exp\left\{-\int_0^\beta d\tau
H_1(\tau)\right\}\rangle_0.
\eeq
In this equation, the symbol T implies an ordering of the operators 
on its right,
from left to right in decreasing order of their time arguments; ${\cal Z}_0=
{\rm tr\;}e^{-\beta H_0}$
and, for any operator
$\cal O$,
\beq\label{operator_ev}
\langle {\cal O} \rangle_0\equiv{\rm Tr}\left({e^{-\beta
H_0}\over {\cal Z}_0}{\cal O}\right)
.\eeq
One commonly  refers to $\tau$ as
the ``imaginary time'' ($\tau$ is real). This  $\tau$ has no
direct physical interpretation: its role here is to properly keep 
track of ordering
of operators in the perturbative expansion.

In field theory, it is often more convenient to use the formalism of path
integrals\index{path integral}.  Let us recall for
instance that for one particle in one dimension the matrix element of the
evolution  operator can be written as
\beq
\langle q_{2}|e^{-iHt}|q_{1}\rangle =\int_{ q\left(
0\right) =q_{1}}^{q\left( t\right) =q_{2}   }{\cal D}\left(
q\left( t\right) \right) \quad e^{i\int_{t_{1}}^{t_{2}}\left( \frac{1}{2}m
\dot{q}^{2}-V\left( q\right)\right) dt } \,,
\eeq
where $q_1$ and $q_2$ denote the positions of the particle at times $0$ and $t$
respectively. 
  Changing $
t\rightarrow -i\tau ,$ and taking the trace, one obtains the following formula
for the partition function:
\beq
{\cal Z}={\rm tr}\;e^{-\beta H}=\int_{q(\beta)=q(0)}{\cal
D}(q)\exp\left\{-\int_0^\beta\left({1\over2}m\dot q^2+V(q)\right)\right\}.
\eeq

  This expression immediately generalizes to the case of a scalar
field, for which the  Lagrangian  is of the form:
\bea\label{lagangian_scalar}
{\cal L} &= & {1\over2}\partial_\mu\phi\partial^\mu\phi-
{m^2\over2}\phi^2-V(\phi) \nonumber\\
&
= &{1\over2}(\partial_0\phi)^2-{1\over2}(\nabla\phi)^2
-{m^2\over2}\phi^2-V(\phi).
\eea
Again, we replace $t$ by
$-i\tau$, $\partial_0=\partial_t$ by $i\partial_\tau$, so
that $(\partial_0\phi)^2\to -(\partial_\tau\phi)^2$. The
partition function  becomes then (integrations over spatial coordinates are
implicit):
\bea\label{Z_scalar}
\lefteqn{ {\cal Z}=\int{\cal D}(\phi)\exp\left\{
-\int_0^\beta
d\tau\left(
{1\over2}(\partial_\tau\phi)^2+{1\over2}(\nabla\phi)^2+
{m^2\over2}\phi^2+V(\phi)
\right)\right\},}\nonumber\\
& &
\eea
where the integral is over periodic fields: $\phi(0)=\phi(\beta)$.

\begin{petitchar}
\renewcommand{\baselinestretch}{.90}\small
\noindent{\bf Remarks.} i) The partition function (\ref{Z_scalar}) 
may be viewed formally as a sum over
classical field configurations in four dimensions, with particular 
boundary conditions in the (imaginary)
time direction.

ii) At high temperature, $\beta\to 0$, the time dependence of the 
fields play no role. The
partition function becomes that of a classical field theory in three 
dimensions:
\beq
{\cal Z}=\int{\cal D}(\phi)\exp\left\{
-\beta\int {\rm d}^3 r
\left({1\over2}(\nabla\phi)^2+
{m^2\over2}\phi^2+V(\phi)
\right)\right\}.
\eeq
Ignoring the time dependence of the fields amounts to take into 
account only the Matsubara frequency
$i\omega_\nu=0$. We shall discuss later explicit  examples of this 
``dimensional reduction''\index{dimensional reduction}.

iii) Note the Euclidean metric in (\ref{Z_scalar}). Since the 
integrand is the exponential of a
negative definite quantity, it is well suited to numerical 
evaluations, using for instance the
lattice technique.

\end{petitchar}

\subsection{Free propagators}

An important feature of the  path integral representation of the partition
function is the boundary conditions to be imposed on the fields over which one
integrates. For the scalar case considered here, the field has to be 
periodic in
imaginary time, with a period
$\beta$.  Similar conditions hold for the fermion fields, which are
antiperiodic in imaginary time, with the same period $\beta$.
It is instructive to see how these periodicity conditions emerge in the operator
formalism, and for this reason we consider now the free propagators, 
first in the
simple case of the non relativistic many body problem. The generalization to
relativistic field is straightforward.

Let us consider a system with unperturbed hamiltonian:
\beq
H_0=\sum_k\epsilon_k \,a^\dagger_k a_k,
\eeq
where $k$ denotes the set of quantum numbers necessary to specify a single
particle state, for instance the three components of the momentum. 
We define time
dependent creation and annihilation operators in the interaction picture:
\bea\label{a_de_tau}
a^\dagger_k(\tau) &\equiv & e^{\tau H_0}a^\dagger_k
  e^{-\tau H_0}=e^{\epsilon_k\tau}a^\dagger_k\nonumber\\
a_k(\tau) &\equiv & e^{\tau H_0}a_k
  e^{-\tau H_0}=e^{-\epsilon_k\tau}a_k
.\eea
The last equalities follow (for example) from the commutation relations:
\beq
[H_0,a_k^\dagger]=\epsilon_k a_k^\dagger\qquad\qquad
[H_0,a_k]=-\epsilon_k a_k
\eeq
which hold for bosons and fermions.
The single particle propagator can then be obtained by a direct calculation:
\bea\label{freeG}
G_k(\tau_1-\tau_2) &=& \langle{\rm
T}a_k(\tau_1)a_k^\dagger(\tau_2)\rangle_0\nonumber\\
&=& e^{-\epsilon_k(\tau_1-\tau_2)}
\left[\theta(\tau_1-\tau_2)(1\pm
n_k)\pm n_k\theta(\tau_2-\tau_1)\right],
\eea
where:
\beq
n_k\equiv \langle a^\dagger_k
a_k\rangle_0=\frac{1}{e^{\beta\epsilon_k}\mp 1}, \eeq
and the upper (lower) sign is for bosons (fermions). 
 One can verify on the expression
(\ref{freeG}) that, in the interval $-\beta<\tau=\tau_1-\tau_2<\beta$,
$G_k(\tau)$ is a periodic (boson) or antiperiodic (fermion) function of
$\tau$:
\beq\label{periodicity}
G_k(\tau-\beta)=\pm G_k(\tau)\qquad\qquad(0\le\tau\le \beta).
\eeq
(To show this relation note that ${\rm e}^{\beta\epsilon_k}n_k=1\pm
n_k$.) It can therefore be represented by a Fourier series
\beq
\label{G_de_tau}
  G_k(\tau)=\frac{1}{\beta}\sum_\nu
e^{-i\omega_\nu\tau}G_k(i\omega_\nu),
\eeq
where the $\omega_\nu$'s are called the Matsubara frequencies\index{Matsubara
frequencies}:
\bea
\begin{array}{llll}
\omega_\nu &=& 2\nu\pi/\beta & \qquad{\rm bosons,} \\
\omega_\nu &=& (2\nu+1)\pi/\beta & \qquad{\rm fermions.}
\end{array}
\eea
The inverse transform is given by
\beq\label{G_de_omega}
G(i\omega_\nu) = \int_0^\beta {\rm d}\tau\, e^{i\omega_\nu\tau}G(\tau)
= \frac{1}{H_0-i\omega_\nu}.
\eeq
Using the property
\beq
\delta(\tau)={1\over\beta}\sum_\nu e^{-i\omega_\nu\tau}\qquad\qquad 
-\beta<\tau<\beta
\eeq
and (\ref{G_de_tau}), it is easily
seen that $G(\tau)$ satisfies the differential equation
\beq\label{equa_diff_G}
(\del_\tau+H_0)\,G(\tau)=\delta(\tau)
,\eeq
which may be also verified directly from (\ref{freeG}).
Alternatively, the single propagator at finite temperature may be obtained as
the solution of this equation with periodic (bosons) or antiperiodic 
(fermions) boundary conditions.

\begin{petitchar}
\renewcommand{\baselinestretch}{.90}\small
\noindent{\bf Remark.} The periodicity or antiperiodicity that we 
have uncovered on the explicit form of
the unperturbed propagator is, in fact, a general property of the 
propagators of a many-body system in
thermal equilibrium. It is a consequence of the commutation relations 
of the creation and annihilation
operators and the cyclic invariance of the trace.
\end{petitchar}

The propagator of the free scalar field
$
\Delta(\tau)=\langle {\rm T}\phi(\tau_1)\phi(\tau_2\rangle,
$
  where $\tau\equiv\tau_1-\tau_2$ satisfies the differential equation
\beq
\left[-\partial_{\tau_1}^2-\nabla_1^2+m^2\right]
\Delta(\tau_1\vec{r_1};\tau_2\vec{r_2})=
\delta(\tau_1-\tau_2)\delta(\vec{r_1}-\vec{r_2}),
\eeq
and obeys periodic boundary conditions. 
It admits the Fourier representation
\beq\label{propscalar}
\Delta(\tau)=\frac{1}{\beta}\sum_n
e^{-i\omega_n\tau}\Delta(i\omega_n) ,
\eeq
where $\omega_n=2\pi n/\beta$ and
\beq\label{scalarmatsu}\Delta(i\omega_n)=\frac{1}{\epsilon_k^2-\omega_n^2}.\eeq
  By inverting the Fourier transform (\ref{scalarmatsu}), one gets
\beq\label{Dtau}
\Delta(\tau)=\frac{1}{2\epsilon_k}\left\{
(1+N_k)e^{-\epsilon_k|\tau|}+N_ke^{\epsilon_k|\tau|}\right\},
\eeq
with $N_k=1/({\rm e}^{\beta\epsilon_k}-1)$.

\subsection{Classical field approximation and dimensional
reduction}\index{dimensional reduction}

In the high temperature limit, $\beta\to 0$, the imaginary-time dependence
of the fields frequently becomes unimportant and can be ignored in a first
approximation. The
integration over imaginary time becomes then trivial  and the
partition function (\ref{Z_scalar})  reduces to:
\beq\label{ZCEcl}
Z\,\approx\,{\cal N}
\int{\cal D}(\phi)\exp\left\{
- \beta \int{\rm d}^3x \,{\cal H}({\bf x})\right\}\,,\eeq
where { $\phi\equiv\phi({\bf x})$ is now a three-dimensional
field, and}
\beq\label{LEcl}
{\cal H}\,=\,{1\over2}\,(\nbfgrad\phi)^2+
{m^2\over2}\,\phi^2 + V(\phi)\,.\eeq
The functional integral { in (\ref{ZCEcl})}
is recognized as the partition function for static  three-dimensional
field configurations with energy $ \int{\rm d}^3x \,{\cal H}(x)$.
We shall refer to this limit as the {\it classical field
approximation}\index{classical field approximation}.

Ignoring the time dependence of the fields is equivalent to retaining only
the zero Matsubara frequency in their Fourier decomposition. Then the Fourier
transform of the free propagator is simply:
\beq
G_0({\bf k})\,=\,\frac{T}{\varepsilon_k^2}\,.
\eeq
This may be obtained directly from  (\ref{propscalar}) keeping only the term
with $\omega_\nu=0$, or from eq.~(\ref{Dtau}) by ignoring
the time dependence and using the approximation
\beq\label{BESOFT}
N(\varepsilon_k)\,=\,\frac{1}{{\rm e}^{\beta\varepsilon_k} -1}\approx
\frac{T}{\varepsilon_k}\,.
\eeq
Both approximations make sense only for $\varepsilon_k\ll T$, implying
$N(\varepsilon_k)\gg 1$. In this limit, the energy per mode is
$\propto\varepsilon_k N(\varepsilon_k)\approx T$,  as expected from the
classical equipartition theorem.

The classical field approximation may be viewed as the leading
term  in a systematic expansion. To see that, let us expand the field
variables in the path integral (\ref{Z_scalar}) in terms of their Fourier
components:
\beq
\phi(\tau)=\frac{1}{\beta} \sum_\nu {\rm e}^{-i\omega_\nu \tau}
\phi(i\omega_\nu),
\eeq
where the $\omega_\nu$'s are the Matsubara frequencies.
The path integral (\ref{Z_scalar}) can then be written as:
\beq\label{ZCE0}
Z\,=\,{\cal N}_1
\int{\cal D}(\phi_0)\exp\left\{-S[\phi_0]\right\}\,,\eeq
where $\phi_0\equiv \phi(\omega_\nu=0)$ depends only on spatial coordinates, and
\beq\label{ZCE1}
\exp\left\{-S[\phi_0]\right\}=\,{\cal N}_2
\int{\cal D}(\phi_{\nu\ne 0})\exp\left\{
- \int_0^\beta {\rm d}\tau \int{\rm d}^3x \,{\cal L}_E(x)\right\}\,.\eeq
The quantity $S[\phi_0]$ may be called the effective action for the ``zero
mode'' $\phi_0$. Aside from the direct classical field contribution
that we have
already considered, this effective action  receives also contributions from
integrating out the non-vanishing Matsubara frequencies\index{Matsubara
frequencies}.
   Diagrammatically, $S[\phi_0]$ is the sum of all the connected
diagrams with external lines associated to
$\phi_0$, and in which the internal lines are the
propagators of the non-static
modes $\phi_{\nu\ne 0}$. Thus, a priori,  $S[\phi_0]$ contains
operators of arbitrarily high order in $\phi_0$, which are
also non-local.
In practice, however, one wishes to expand  $S[\phi_0]$ in terms  of
{\it local} operators, i.e., operators with the schematic structure
$a_{m,\,n}\grad^m \phi_0^n$ with coefficients $a_{m,\,n}$
to be computed in perturbation theory.

To implement this strategy, it is useful to introduce an intermediate scale
$\Lambda$ ($\Lambda\ll  T$) which  separates {\it hard}
($k\simge \Lambda$)
and {\it soft} ($k\simle \Lambda$) momenta. All the non-static modes, as well
as the static ones with $k \simge \Lambda$ are {\it hard}
(since $K^2\equiv \omega_\nu^2 +k^2 \simge \Lambda^2$
for these modes), while the static ($\omega_\nu=0$) modes
with $k\simle \Lambda$ are {\it soft}.
Thus, strictly speaking, in the construction of the effective
theory along the lines indicated above, one has to integrate out also
the static
modes with $k \simge\Lambda$. The benefits of this separation of
scales are that ({\it a}) the resulting effective action for the
soft fields can be made
   {\it local} (since the initially non-local amplitudes can be
expanded out in powers of $p/K$, where $p \ll \Lambda$
is a typical external momentum, and $K\simge
\Lambda$ is a hard momentum on an
internal line), and  ({\it b}) the effective theory is now
used exclusively at soft momenta, where  classical approximations such as
(\ref{BESOFT}) are expected to be valid.
This strategy, which consists in integrating out the non-static modes
   in perturbation theory in order to obtain an effective
three-dimensional  theory for the soft static modes (with
$\omega_\nu=0$ and $k\equiv |{\bf k}|\simle \Lambda$), is generally
referred  to as ``dimensional reduction''
\cite{Gins80,Appel81,Nadkarni83,Landsman89,Braaten94,Kajantie94}\index{dimensional
reduction}.

   As an illustration
let us consider a massless scalar theory with quartic interactions;
that is, $m=0$ and $V(\phi)=({g^2}/{4!})\phi^4$ in
(\ref{lagangian_scalar}). The ensuing effective action\index{effective action}
for the soft fields (which we shall still denote as $\phi_0$) reads
\bea\label{EFFLagran}
& & S[\phi_0] =\beta {\cal F}(\Lambda)\nonumber \\
& & +\int{\rm d}^3x \,\left\{
{1\over 2}\,(\nbfgrad\phi_0)^2+{1\over 2}\,M^2(\Lambda)\phi^2_0+
\,{g^2_3(\Lambda)\over 4!}\,\phi^4_0+{h(\Lambda)\over 6!}\,\phi^6_0
+\Delta {\cal L}\right\},\nonumber\\  & &
\eea
where ${\cal F}(\Lambda)$ is the contribution of the hard modes
to the free-energy, and
$\Delta {\cal L}$ contains all the other local operators which are
invariant under rotations and under the  symmetry $\phi\to -\phi$,
i.e., all the
local operators which are consistent with the symmetries of the
original Lagrangian.
We have changed the normalization of the field ($\phi_0\rightarrow
\sqrt{T}\phi_0$) with respect to  (\ref{ZCEcl})--(\ref{LEcl}),
so as to
absorb the factor $\beta$ in front of the effective action. The
effective ``coupling
constants'' in  (\ref{EFFLagran}), i.e.
$M^2(\Lambda)$,
$g^2_3(\Lambda)$,
$h(\Lambda)$ and the infinitely many parameters in $\Delta {\cal L}$,
are computed
in perturbation theory, and depend upon the
separation scale $\Lambda$, the temperature $T$ and the original
coupling $g^2$. To lowest order in $g$, $g^2_3\approx g^2T$,
$h\approx 0$ (the first contribution to $h$
arises at order $g^6$, via one-loop diagrams), and $M\sim gT$,
as we shall see shortly.
Note that eq.~(\ref{EFFLagran}) involves in general non-renormalizable
operators, via $\Delta {\cal L}$. This is not a difficulty, however,
since this is only an { effective} theory, in which the scale
$\Lambda$ acts as an explicit ultraviolet (UV) cutoff for
the loop integrals. Since however the scale $\Lambda$ is
arbitrary, the dependence on $\Lambda$ coming from such soft loops
must cancel against the dependence on $\Lambda$ of the parameters
in the effective action.

\begin{figure}
\includegraphics[width=.1\textwidth]{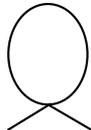}
	 \caption{One-loop tadpole diagram for the self-energy of the 
scalar field}
\label{tadpole}
\end{figure}
Let us verify this cancellation explicitly
in the case of the thermal mass $M$ of the scalar field,
and to lowest order in perturbation theory. To this order,
the scalar self-energy is given by the tadpole diagram in
Fig.~\ref{tadpole}. The
mass parameter $M^2(\Lambda)$ in the effective action is obtained by
integrating
over hard momenta within the loop in
   Fig.~\ref{tadpole}:
\bea\label{MLAM}
M^2(\Lambda)&=&{g^2\over 2}\,T\sum_\nu
\int\frac{{\rm d}^3k}{(2\pi)^3}\, \frac{(1-\delta_{\nu 0})+
\theta(k-\Lambda)\delta_{\nu 0}}{\omega_\nu^2+k^2}\nonumber\\
&=&{g^2\over 2}\int\frac{{\rm d}^3k}{(2\pi)^3}\,
\left\{\frac{N(k)}{k}\,+\,\frac{1}{2k}\,-\,\theta(\Lambda-k)
\frac{T}{k^2}\right\},\eea
where the $\theta$-function in the second line has been
generated by writing  $\theta(k-\Lambda) = 1 -\theta(\Lambda-k)$.
The first term, involving the thermal distribution, gives
the contribution
\beq\label{MSHTL}
\hat M^2\,\equiv\,
\frac{g^2}{2}\int\frac{{\rm d}^3k}{(2\pi)^3}\,\frac{N(k)}{k}
\,=\,\frac{g^2}{24}\,T^2\,.\eeq
As it will turn out, this is the leading-order (LO)
scalar thermal mass, and also the simplest example of what will be called
  ``hard thermal loops''
(HTL). The second term, involving $1/2k$, in (\ref{MLAM}) is
quadratically UV divergent, but independent of the temperature;
   the standard
renormalization procedure at $T=0$
amounts to simply removing this term.
The third term, involving the $\theta$-function, is easily
evaluated. One finally gets:
\beq\label{MLAM1}
M^2(\Lambda)\,=\,\hat M^2\,-\,\frac{g^2}{4\pi^2}\,\Lambda T\,
\equiv\,
\frac{g^2T^2}{24}\left(1-\frac{6}{\pi^2}\,\frac
{\Lambda}{T}\right).\eeq
The $\Lambda$-dependent term above is subleading, by a
factor $\Lambda/T\ll 1$.

The one-loop correction to the thermal mass within the effective
theory is given by the same diagram in Fig.~\ref{tadpole}, but
where the internal field is static and soft, with the massive
propagator $1/(k^2+M^2(\Lambda))$,
and coupling constant $g^2_3\approx g^2T$. Since the typical momenta in the
integral will be $k\simge M$, and $M\sim\hat M\sim gT$, we choose $\Lambda\gg
gT$. We then obtain
\bea\label{MDCL1}
\delta M^2(\Lambda) &=&
{g^2\over 2}\int\frac{{\rm d}^3k}{(2\pi)^3}\,\Theta(\Lambda-k)\,
\frac{T}{k^2+M^2(\Lambda)}\nonumber\\
&=&\frac{g^2T\Lambda}{4\pi^2}\left(
1-{\pi M\over 2\Lambda}\,\arctan\,{\Lambda\over M }\right)
\,\simeq\,
\frac{g^2T\Lambda}{4\pi^2}
\,-\,\frac{g^2}{8\pi}\hat MT\,,\eea
where the terms neglected in the last step
are of higher order in $\hat M/\Lambda$
or $\Lambda/T$.

As anticipated, the $\Lambda$-dependent terms cancel in the
sum $M^2\equiv M^2(\Lambda)+\delta M^2(\Lambda)$, which then provides
the physical thermal mass within the present accuracy:
   \beq \label{MTOTS}
M^2\,=\,M^2(\Lambda)+\delta M^2(\Lambda)
\,=\,\frac{g^2T^2}{24}
\,-\,\frac{g^2}{8\pi}\hat MT\,.\eeq
The LO term, of order $g^2T^2$, is the HTL $\hat M$.
The next-to-leading order (NLO) term, which involves
the resummation of the thermal mass $M(\Lambda)$ in the
soft propagator, is of order $g^2\hat MT\sim g^3T^2$,
and therefore non-analytic in $g^2$.
This non-analyticity is related to the fact that the integrand
in  (\ref{MDCL1}) cannot be expanded in powers of $M^2/k^2$
without generating infrared divergences.

\section{Effective theories for the quark-gluon plasma }\index{effective
theory}

We return now to the quark-gluon plasma and analyze the various   
scales and degrees of freedom\index{degrees of freedom} which are relevant in
the weak coupling regime. We show that there is a hierarchy of scales
controlled by powers of the gauge coupling
$g$. We
 focus in these lectures on two particular momentum scales, the \lq hard'
one which is that of the plasma particles with momenta $k \sim T$, and the \lq
soft' one with
$k\sim gT$ at which collective phenomena develop. We shall be in particular
interested in the effective theory obtained when the hard degrees of freedom
are \lq integrated out'. The resulting effective theory describe long
wavelength, low frequency collective phenomena; that is, it accounts for time
dependent fields, in contrast to the example discussed in the previous section
which concerned only static fields. As we shall see later, getting a complete
description of the dynamics of the collective excitations turns out to be
important also for the calculation of the equilibrium  properties
of the quark-gluon plasma.

\subsection{Scales and degrees of freedom in ultrarelativistic plasmas}

A property of QCD which is essential in the present discussion is that of
asymptotic freedom, according to which the coupling constant depends on the
scale $\bar\mu$ as 
\beq
\alpha_s(\bar \mu)\,\equiv\,{g^2\over 4\pi}\,\propto\,
       \frac{1}{\ln(\bar \mu/\Lambda_{QCD})}.
\eeq
At high temperature, the natural scale  is 
$\bar \mu \,=\,2\pi T$, so that the coupling becomes weak when
$2\pi T\gg\Lambda_{QCD}$.  At
extremely high temperature the interactions become negligible and
hadronic matter turns into an ideal gas of quarks and gluons: this is
the quark-gluon plasma. As we shall see an important effect of the
interactions is to turn free quarks and gluons into
weakly interacting quasiparticles. 

In the absence of interactions, the plasma particles\index{plasma particles}
are distributed in momentum space
according to the Bose-Einstein  or  Fermi-Dirac distributions:
\beq\label{BFINTRO}
N_k\,=\,\frac{1}{{\rm e}^{\beta \varepsilon_k}\,-\,1},\qquad\qquad
n_k\,=\,\frac{1}{{\rm e}^{\beta \varepsilon_k}\,+\,1},\eeq
where $\varepsilon_k=k\equiv |{\bf k}|$ (massless particles),
$\beta\equiv 1/T$, and chemical potentials are assumed to vanish. In such a
system, the particle density
$n$ is
   determined by the temperature: $n\propto T^3$. Accordingly,
the mean interparticle distance $n^{-1/3}\sim 1/T$ is of the same
order as the thermal wavelength $\lambda_T=1/k$ of  a
typical particle in the thermal bath for which $k\sim T$. Thus the particles of
an ultrarelativistic plasma are  quantum degrees of freedom for which in
particular  the Pauli principle can never be ignored.

In the weak coupling regime ($g\ll 1$), the interactions do not alter
significantly the picture. The hard degrees of freedom, i.e. the plasma
particles, remain the dominant degrees of freedom and since the coupling to
gauge fields occurs typically through covariant derivatives,
$D_x=\del_x+igA(x)$,  the effect of interactions on particle motion is a small
perturbation unless the fields are very large, i.e., unless
$A\sim T/g$, where $g$ is the gauge coupling: only then do we have
$\del_X\sim T\sim gA$, where $\del_X$ is a space-time gradient.
We should note here that   we
  rely on considerations,  based
on the magnitude of the gauge fields, which
depend on the
choice of a gauge. What is meant is  that  there exists  a large
class of gauge choices for which they are valid. And we
shall verify a posteriori that within such a class, the final results are
gauge invariant.

Considering now more generally the effects of the interactions, we note that
these depend both on the strength of the gauge fields and on the wavelength of
the modes under study. A measure of the strength of the gauge fields in
typical situations is obtained from the magnitude of their thermal
fluctuations\index{thermal fluctuations}, that is
$\bar A\equiv \sqrt{\langle A^2(t,{\bf x})\rangle}$. In equilibrium $\langle
A^2(t,{\bf
x})\rangle$ is independent of $t$ and ${\bf x}$ and given by
$\langle A^2\rangle= G(t=0,{\bf x=0})$
where $G(t,{\bf x})$ is the gauge field propagator.
In the non interacting case  we have (with $\varepsilon_k=k$):
\beq\label{fluctuationsA}
\langle A^2\rangle=
\int \frac{{\rm d}^3
k}{(2\pi)^3}\frac{1}{2\varepsilon_k}(1+2N_k).
\eeq
Here we shall use this formula also in the interacting case,
assuming that  the  effects of the interactions can be accounted for  simply
by a  change of $\varepsilon_k$.
   We shall also ignore
   the (divergent) contribution of
   the vacuum fluctuations (the term independent of the temperature
   in (\ref{fluctuationsA})).

   For
the plasma particles   $\varepsilon_k=k\sim T$ and  $\langle 
A^2\rangle_T\sim T^2$.
The associated electric (or magnetic) field fluctuations are
$\langle E^2 \rangle_T\sim
\langle (\del A)^2\rangle_T \sim k^2 \langle A^2\rangle_T\sim
T^4
$ and are a dominant contribution to the plasma energy density.
  As already mentioned, these short wavelength,
or {\it hard}, gauge field
fluctuations produce a small perturbation on the motion of a plasma particle.
However, this is not so for an excitation at the momentum scale  $k\sim gT$,
since then  the two terms in the covariant derivative
$\del_X$ and $g\bar A_T$ become comparable. That is, the properties of an
excitation with momentum $gT$ are expected to be non perturbatively 
renormalized
by the hard thermal fluctuations. And indeed, the scale
$gT$ is that  at which collective phenomena develop. The emergence of the Debye
screening mass
$m_D\sim gT$ is one of the simplest examples of such phenomena.

Let us now consider the fluctuations at this scale
$gT\ll T$, to be referred to as the {\it soft} scale.
These fluctuations can be accurately described by classical
fields. In fact the associated occupation numbers
$N_k$ are large, and
accordingly one can replace
$N_k$ by $ T/\varepsilon_k$ in (\ref{fluctuationsA}).
Introducing  an upper cut-off $gT$ in the momentum integral, one  then
gets:
\beq
\langle A^2\rangle_{gT} \sim \int^{gT}{\rm d}^3k \, \frac{T}{k^2}\sim
gT^2.
\eeq
Thus $\bar A_{gT}\sim \sqrt{g} T$ so that $g\bar A_{gT}
\sim g^{3/2}T$ is still of higher order than the kinetic term $\del_X\sim gT$.
In that sense the soft modes with $k\sim gT$ are still perturbative, i.e. their
self-interactions can be ignored in a first approximation. Note however that
they generate contributions to physical observables which are not analytic in
$g^2$, as shown by the example of the order
$g^3$ contribution to the
energy density of the plasma:
\beq
\epsilon^{(3)}\sim  \int_0^{\omega_{pl}}
{\rm d}^3 k \,\,\omega_{pl}\,\frac{1}{{\rm e}^{\omega_{pl}/T}-1}\sim
\omega_{pl}^3\,\omega_{pl}\,\frac{T}{\omega_{pl}}\,\sim\, g^3T^4,
\eeq
where $\omega_{pl}\sim gT$ is the typical frequency of a collective mode.

Moving down to a lower momentum scale, one meets the contribution of the
unscreened magnetic fluctuations which play a dominant role for
$k\sim g^2T$.  At that scale, to be referred to as the {\it ultrasoft }
scale,  it becomes necessary to distinguish  the electric and the magnetic
sectors (which provide comparable contributions at the scale
$gT$). The  electric
fluctuations are damped by the Debye screening mass
($\varepsilon_k^2=k^2+m_D^2\approx m_D^2$ when $k\sim g^2T$)
  and their contribution is  negligible, of order  $ g^4
T^2$. However, because of
the absence of static screening in the magnetic sector, we have here
$\varepsilon_k\sim k$ and
\beq\label{fluctg2t}
\langle A^2\rangle_{g^2T}\sim T\int_0^{g^2T}{\rm d}^3k \frac{1}{k^2}\,
\sim\,g^2 T^2,\eeq
so that $g\bar A_{g^2T}\sim g^2T$ is now of the same order as the
  ultrasoft
derivative $\del_X\sim g^2T$: the fluctuations are no longer perturbative. This
is the origin of the breakdown of perturbation theory in high temperature QCD.

\begin{figure}
\includegraphics[width=.5\textwidth]{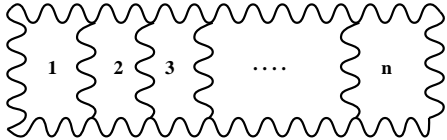}
\label{fig:ladders}
\caption{Example of a multiloop diagram which is infrared divergent}
\end{figure}

To appreciate the difficulty from another perspective,
let us first observe that the dominant contribution to
the fluctuations at scale $g^2T$ comes from the zero Matsubara frequency:
\beq
\langle A^2\rangle_{g^2T}= T\sum_n \int_0^{g^2T}{\rm d}^3k\,\,
\frac{1}{\omega_n^2+k^2}\sim T\int_0^{g^2T}{\rm d}^3k \,\,\frac{1}{k^2}.
\eeq
Thus the fluctuations that we are discussing are those of a three
dimensional theory of static fields. Following
Linde \cite{Linde79,Linde80}  consider then the higher order
corrections to the pressure in hot  Yang-Mills theory. Because of the strong
static fluctuations most of the diagrams of perturbation theory are 
infrared (IR)
divergent. By power counting\index{power counting}, the strongest IR
divergences arise from ladder diagrams, like the one depicted in
Fig.~\ref{fig:ladders}, in which  all the propagators are static, and the loop
integrations are three-dimensional. Such
$n$-loop diagrams can be estimated as ($\mu$ is an IR cutoff):
\beq\label{Linde}
g^{2(n-1)}\left(T\int {\rm d}^3k\right)^n\frac{k^{2(n-1)}}
{(k^2+\mu^2)^{3(n-1)}}\,,\eeq
which is of the order $g^6 T^4 \ln(T/\mu)$ if $n=4$ and of the order
$g^6T^4\left(g^2T/\mu\right)^{n-4}$  if $n>4$. (The various factors in
(\ref{Linde}) arise, respectively, from the $2(n-1)$ three-gluon vertices,
the $n$ loop integrations, and the $3(n-1)$ propagators.) According
to this equation, if $\mu \sim g^2T$,
all the diagrams with
$n\ge 4$ loops contribute to the same order, namely to ${\cal O}(g^6)$.
In other words, the correction of ${\cal O}(g^6)$ to the pressure cannot be
computed in perturbation theory.

\subsection{Effective theory at scale $gT$}

Having identified the main scales and degrees of freedom\index{degrees of
fredom}, our  task will be to construct appropriate effective
theories\index{effective theory} at the various scales, obtained by eliminating
the degrees of freedom at higher scales. We shall  consider here the effective
theory at the scale
$g$T obtained by eliminating the hard degrees of freedom with momenta $k\sim
T$.

The soft
excitations at the scale $gT$ can be described in terms of
{\it average  fields} \cite{qed,qcd}. Such average fields develop for example
when the system is exposed to an external perturbation, such as an external
electromagnetic current. In QED, we can summarize the  effective theory for the
soft modes by the equations of motion:
\beq\label{maxwell00}
\del_\mu F^{\mu\nu}=j_{ind}^\nu+j_{ext}^\nu
\eeq
that is, Maxwell equations with a source term composed of the external
perturbation $j_{ext}^\nu$, and an extra contribution $j_{ind}^\nu$ which we
shall refer to as the {\it induced current}. The induced current is generated
by  the collective motion of the charged particles, i.e. the hard degrees of
freedom. It may be regarded itself as a functional of the average gauge  fields
and, once this functional is known,  the equations above constitute a closed
system of equations for the soft fields.

The main problem is to
calculate
$j_{ind}$. This is done by considering the dynamics of the hard
particles in the
background of the soft fields. For QED, the induced current  can be obtained
using linear response theory.
   To be more specific, consider as an
example a system of charged particles on which is acting a perturbation
of the form  $\int
{\rm d}x \,j_\mu (x)
A^\mu (x)$, where $j_\mu(x)$ is the current operator and $A^\mu (x)$ some
applied gauge  potential. Linear response theory leads to the
following relation
for the induced current:
\bea\label{responsej}
& &j_\mu^{ind}=\int {\rm d}^4 y\,  \Pi^R_{\mu\nu}(x-y) A^\nu (y), \nonumber\\
& & \qquad\qquad\qquad
\Pi^R_{\mu\nu}(x-y)=-i\theta(x_0-y_0)\langle
[j_\mu(x),j_\nu (y)]\rangle_{eq.},
\eea
where the (retarded) response function $\Pi^R_{\mu\nu}(x-y)$ is also 
referred to
as the polarization operator. Note that in (\ref{responsej}), the
expectation value is  taken in the equilibrium state.  Thus, within
linear response, the task of calculating the basic  ingredients of 
the effective
theory for soft modes  reduces to  that of calculating appropriate
equilibrium correlation functions.

  In fact we shall need the
response function only in the
weak coupling regime, and for particular kinematic conditions
which allow for
important simplifications. In leading order in weak coupling, the
polarization tensor is given by the one-loop approximation. In the kinematic
regime of interest, where the incoming momentum is soft while the loop momentum
is hard, we can write
$\Pi(\omega,p)=g^2T^2f(\omega/p,p/T)$ with $f$ a dimensionless function, and in
leading order in
$p/T\sim g$,
$\Pi$ is of the form $g^2T^2f(\omega/p)$. This particular contribution of the
one-loop polarization tensor is an example of what has been called a ``hard
thermal loop''
\cite{Klimov81,Weldon82a,Weldon82b,Pisarski89,FT90,Nair91,qed,qcd}\index{hard
thermal loop}; for photons in QED, this is the only one. It turns out that
this  hard thermal loop can be obtained from simple {\it kinetic theory}, and
the corresponding calculation is done in the next subsection.

In non Abelian theory,
linear response is not sufficient:
constraints due to gauge symmetry force us to take into account
specific non linear effects and a more complicated formalism needs to
be worked out. Still,  simple kinetic equations can  be
obtained in this case also, but in contrast to QED, the resulting induced
current is a non linear  functional of the gauge fields. As a result, it
generates  an infinite number of  ``hard thermal loops''.

\section{Kinetic equations for the plasma particles}

The hard degrees of freedom enter the equations of motion (\ref{maxwell00}) for
the soft collective excitations only through their average density or current,
that is, through the induced current. This induced current can be calculated by
studying the dynamics of the plasma particles in the background of soft
external gauge fields. This is what we now turn to. In order to keep the
discussion at an elementary level, we shall merely analyze  the main steps 
involved in the derivation of the corresponding QCD equations in the simpler
context of non relativistic electromagnetic plasmas.  The QCD equations are
presented at the end of this section. 

\subsection{One-loop polarization tensor from kinetic theory}

As indicated above, in the kinematic regime
considered, the dominant contribution to the one loop polarization tensor can
be obtained using elementary kinetic theory, and we present now this
 calculation.
   We consider an electromagnetic plasma and  momentarily assume that we
can describe its charged particles in terms of  classical distribution
functions
$f_q({\bf p},x)$ giving the density of particles of charge  $q$ ($q=\pm e$)
and
momentum ${\bf p}$ at the space-time point $x=(t,{\bf r})$
\cite{PhysKin}.
We consider then the case where collisions among the charged particles can be
neglected and where the only relevant interactions are those of particles with
average electric (${\bf E}$) and magnetic (${\bf B}$) fields. Then the
distribution functions obey the following simple kinetic equation, known as
the Vlasov equation \cite{PhysKin}\index{Vlasov equation}:
\beq\label{vlasov}
\frac{\del f_q}{\del t}+{\bf v}\frac{\del f_q}{\del {\bf r}}+{\bf F}\frac{\del
f_q}{\del {\bf p}}=0,
\eeq
where ${\bf v}={\rm d}\varepsilon_p/{\rm d}{\bf p}$ is the velocity of a
particle with momentum ${\bf p}$ and energy $\varepsilon_p$ (for massless
particles
${\bf v}=\hat{\bf p}$), and
${\bf F}=q({\bf E}+{\bf v}\wedge{\bf B})$ is the Lorentz force.
   The average fields ${\bf E}$ and ${\bf
B}$ depend themselves on the distribution functions $f_q$. Indeed, the induced
current
\beq\label{indcurrent1}
j^\mu_{ind}(x)=e\int \frac{{\rm d}^3p}{(2\pi)^3}\,v^\mu\,\left(f_+({\bf
p},x)-f_-({\bf p},x)\right),
\eeq
where  $v^\mu\equiv (1,{\bf v})$, is the source term in
the Maxwell equations (\ref{maxwell00}) for the mean fields.

   When the
plasma is in equilibrium, the distribution functions, denoted  as
$f_q^0(p)\equiv f^0(\varepsilon_p)$, are
isotropic in momentum space and independent of space-time coordinates; the
induced current vanishes, and so do the average fields
${\bf E}$ and
${\bf B}$. When the plasma is weakly perturbed, the distribution functions
deviate slightly from their equilibrium values, and we can write:
$f_q({\bf p},x)=f^0(\varepsilon_p)+\delta f_q({\bf p},x)$.
In the linear approximation,  $\delta f$ obeys
\beq\label{vlasovlinear}
(v\cdot\del_x)\delta f_q({\bf p},x)=-
q{\bf v}\cdot {\bf E}\frac{{\rm d}f^0}{{\rm
d}\varepsilon_p},
\eeq
where $v\cdot\del_x\equiv \del_t+{\bf v}\cdot {\bf \grad}$. The magnetic field
does not contribute because of the isotropy of the
equilibrium distribution function.

It is convenient here to set
\beq\label{DEFW}
\delta f_q({\bf p},x)\,\equiv\,-qW(x,{\bf v})\,\frac{{\rm d}f^0}{{\rm
d}\varepsilon_p},
\eeq
thereby introducing a notation which will be useful later for 
the QCD case.
Since
\beq
f_q({\bf p},x)\,=\,f^0(\varepsilon_p)
-qW(x,{\bf v})\,\frac{{\rm d}f^0}{{\rm
d}\varepsilon_p}\,\simeq\, f^0(\varepsilon_p-qW(x,{\bf v})),\eeq
$W(x,{\bf v})$ may be
viewed as a local distortion of the momentum distribution of the
plasma particles. The equation for $W$ is simply:
\beq\label{EQW}
(v\cdot\del_x)W(x,{\bf v})\,=\,{\bf v}\cdot {\bf E}(x).\eeq

Contrary to (\ref{vlasov}),
the linearized equations (\ref{vlasovlinear}) or (\ref{EQW})
do not involve the derivative of $f$ with respect to
${\bf p}$, and they can be solved by the method of characteristics:
$v\cdot\del_x$
is the time derivative of
$\delta f({\bf p},x)$ along the characteristic defined by ${\rm d }{\bf x}/{\rm
d}t={\bf v}$. Assuming then that
the perturbation is introduced adiabatically so that the fields and the
fluctuations vanish as ${\rm e}^{\eta t_0}$ ($\eta\to 0^+$) when
$t_0\to -\infty$, we obtain the retarded solution:
\beq\label{fluctuation}
W(x,{\bf v})\,=\,
\int_{-\infty}^t{\rm d}t'\,{\bf v}\cdot {\bf E}({\bf
x-v}(t-t'),t'),
\eeq
and the corresponding  induced current:
\beq\label{JMU0}
j^\mu_{ind}(x)= -2e^2\int \frac{{\rm d}^3p}{(2\pi)^3}\,v^\mu\frac{{\rm
d}f^0}{{\rm d}\varepsilon_p}\int_0^\infty{\rm d}\tau\, {\bf v}\cdot{\rm
E}(x-v\tau).
\eeq
Since ${\bf E}=-{\bf \grad}A^0-\del {\bf
A}/\del t$, the induced current is a linear functional of
$A^\mu$.
At this point we assume explicitly that the particles are
massless. In this case, ${\bf v}$ is a unit vector, and the angular
integral over the direction of ${\bf v}$ factorizes in (\ref{JMU0}).
Then, using  (\ref{responsej}) as definition for the polarization tensor
$\Pi^{\mu\nu}(x-y)$, and the
fact that the Fourier transform of $\int_0^\infty {\rm d}\tau\, {\rm
e}^{-\eta\tau}f(x-v\tau)$ is $i\,f(Q)/(v\cdot Q+i\eta)$, with
$Q^\mu=(\omega,{\bf q})$ and $f(Q)$ the Fourier transform of $f(x)$, one
gets, after a simple calculation \cite{Silin60} :
\beq\label{polarisation2}
\Pi_{\mu\nu}(\omega,{\bf q})=m_D^2\left\{
-\delta_{\mu 0}\delta_{\nu 0}+\omega\int\frac{{\rm d}\Omega}{4\pi} \frac{ v_\mu
v_\nu}{\omega-{\bf v}\cdot{\bf q}+i\eta}\right\},
\eeq
where the angular integral $\int {\rm d}\Omega$
   runs over all the orientations of ${\bf v}$, and
$m_D$ is the Debye screening mass:
\beq\label{plasmafrequency}
m_D^2 = -\frac{2e^2}{\pi^2}\int_0^\infty{\rm d}p\, p^2
\frac{{\rm d}f^0}{{\rm d}\varepsilon_p}\,.
\eeq
It turns out that (\ref{polarisation2})  is  the dominant contribution at
high temperature to the one-loop polarization tensor in QED,
provided one substitutes
for $f^0$ the actual quantum equilibrium distribution function,
that is, $f^0 (\varepsilon_p)=n_p$, with $n_p$ given in (\ref{BFINTRO}).
After insertion in (\ref{plasmafrequency}), this yields $m_D^2=e^2T^2/3$.

In
the next subsection, we shall address the question of how simple kinetic
equations emerge in the description of systems of quantum particles, and under
which conditions such systems can be described by seemingly classical
distribution functions where both positions and momenta are simultaneously
specified.

We shall later find that the expression obtained for the polarization tensor
using simple kinetic theory generalizes to the non Abelian case.
This is so in
particular because  the kinematic regime  remains that of the linear Vlasov
equation, with  straight line characteristics\index{Vlasov equation}.

\subsection{Kinetic equations for quantum particles}

In order to discuss in a simple setting how  kinetic equations emerge in the
description of collective motions of quantum particles, we consider in this
subsection a system of non relativistic fermions coupled to classical gauge
fields. Since we are dealing with a system of independent particles in
an external field,  all the information on the quantum many-body state  is
encoded in the one-body density matrix\index{density matrix}
\cite{Fetter71,BR86} :
\beq\label{spdm0}
\rho({\bf r},{\bf r'},t)=\langle \Psi^\dagger({\bf r'},t)\Psi({\bf
r},t)\rangle\,,
\eeq
where $\Psi$ and $\Psi^\dagger$ are the annihilation and creation
operators, and
the average is over the initial equilibrium state. It is on this object that we
shall  later implement the relevant kinematic approximations.  To
this aim, we introduce the {\it Wigner transform} of
$\rho({\bf r}, {\bf r}',t)$ \cite{Wigner1,Wigner2}\index{Wigner transform}:
\beq
f({\bf p}, {\bf R},t)=\int {\rm d}^3 s \, {\rm e}^{-i {\bf p}\cdot{\bf s}}\,
\rho\left({\bf R}+\frac{{\bf s}}{2}, {\bf R}-\frac{{\bf
s}}{2},t\right).
\eeq
The Wigner function has many properties that one expects of a classical
phase space distribution function as may be seen by calculating the
expectation values of simple one-body observables. For instance the
average density of particles $n({\bf R})$ is given by:
\beq
n({\bf R}, t)=\rho({\bf R}, {\bf R}, t)=\int\frac{{\rm d}^3 p}{(2\pi)^3}
\,f({\bf p}, {\bf R}, t).
\eeq
Similarly, the current operator:
$({1}/{2mi})\left( \psi^\dagger\bfnabla\psi- (\bfnabla
\psi^\dagger)\psi\right)$
has for expectation value:
\beq\label{jnonrel}
{\bf j}({\bf R}, t)=\frac{1}{2mi}\left(\bfnabla_y-\bfnabla_x\right)\rho({\bf
y},{\bf x}, t)|_{|{\bf y}-{\bf x}|\to 0}=
   \int\frac{{\rm d}^3 p}{(2\pi)^3}\,\frac{{\bf p}}{m}\,
f({\bf p}, {\bf R}, t).
\eeq
These results are indeed those one would obtain in a classical description
with $f({\bf p},{\bf R},t)$ the probability density to find a particle with
momentum ${\bf p}$ at point ${\bf R}$ and time $t$.  Note however that while
$f$ is real, due to the hermiticity of
$\rho$,  it is not always  positive as a truly classical distribution
function would be. Of course $f$ contains the same
quantum information as
$\rho$, and it does  not make sense to specify quantum
mechanically both the position and the momentum. However,  $f$ behaves as
   a classical distribution function in the  calculation of one-body
observables  for which the typical momenta $p$ that are involved in the
integration are large in comparison with the scale $1/\lambda$ characterizing
the range of spatial variations of $f$, i.e. $p\lambda\gg 1$.

     By using the equations of motion for the
field operators,
$i\dot\Psi({\bf r},t)=[H,\Psi]$, where $H$ is the single particle Hamiltonian,
one obtains easily the following
   equation of
motion for the density matrix
\beq \label{hrho}
i \del_t
\rho=[H,\rho].
\eeq
   In fact we shall need the Wigner transform of this equation  in
cases where  the gradients with respect to
$R$ are small compared to the typical values of $p$. Under such conditions, the
   equation of motion
   reduces to
\beq\label{kineteq0}
\frac{\del}{\del t}f+\bfnabla_p \,H\cdot \bfnabla_R \,f-\bfnabla_R \,H\cdot
\bfnabla_p \,f=0.
\eeq
where we have kept only the leading terms in an expansion in $\bfnabla_R $. For
particles interacting with gauge potentials
$A^\mu(X)$, the Wigner transform of the single particle Hamiltonian in
(\ref{kineteq0}) takes  the form:
\beq
H({\bf R},{\bf p},t)=\frac{{\bf p}^2}{2m}-\frac{e}{m}{\bf A}\cdot{\bf p}
+\frac{e^2}{m} {\bf A}^2({\bf R},t)+eA_0({\bf R},t).
\eeq
Assuming that the field is weak and neglecting the term in $A^2$,
one can write (\ref{kineteq0}) in the form:
\beq\label{vlasovnoco}
\del_t f+{\bf v}\cdot\bfnabla_R f+e ({\bf E}+{\bf v}\wedge{\bf
B})\cdot\bfnabla_p f +\frac{e}{m}(p_j\del_j
A^i)\nabla^i_p f=0,
\eeq
where we have set ${\bf v}=({\bf p-eA})/{m}$. This equation is almost
   the Vlasov equation (\ref{vlasov}): it differs
from it by the last term which is not gauge
invariant. The presence of such a term, and the related gauge dependence of
the  Wigner function, obscure the physical interpretation. It is then 
convenient
to  define a gauge invariant density matrix:
\beq\label{rhocov}
\acute\rho({\bf r},{\bf r'},t)=\langle\psi^\dagger({\bf r'},t)\psi({\bf
r},t)\rangle U({\bf r},{\bf r'}, t),
\eeq
where (${\bf s}={\bf r}-{\bf r}'$)
\beq\label{approxU}
U({\bf r},{\bf r'})=\exp\left(-ie\int_{\bf r'}^{\bf r} {\rm d}{\bf z}
\cdot {\bf A}({\bf z}, t))\right)\approx
\exp\left(-ie{\bf s}\cdot {\bf A}({\bf R})\right)
\eeq
and the integral is along an arbitrary path
  going from ${\bf r}'$ to ${\bf r}$. Actually, in the
last step we have used an approximation which amounts to chose for this path
the straight line between ${\bf r}'$ to ${\bf r}$; furthermore, we have assumed
that the gauge potential does not vary significantly between ${\bf 
r}'$ to ${\bf
r}$.  (Typically, $\rho({\bf r},{\bf r}')$ is peaked at $s=0$ and drops to zero
when $s\simge \lambda_T$ where $\lambda_T$ is the thermal wavelength of the
particles. What we assume is that over a distance of order 
$\lambda_T$ the gauge
potential remains approximately constant.)  Note that in the calculation of
the current, only the limit
$s\to 0$ is required, and that is given correctly by (\ref{approxU}) (see
also (\ref{jcov}) below). With the approximate expression (\ref{approxU})
the Wigner transform of (\ref{rhocov}) is simply
$\acute f({\bf R},{\bf k})=f({\bf R},{\bf k} +e{\bf A})$. By making the
substitution
$f({\bf R},{\bf p})=\acute f({\bf R},{\bf p}-e{\bf A})$ in
(\ref{vlasovnoco}), one verifies that the  non covariant
term cancels out and that the covariant Wigner function $\acute f$ obeys indeed
   Vlasov's equation.

In the presence of a gauge field, the previous definition (\ref{jnonrel})
of the current suffers
from  the lack of gauge covariance. It is however easy to construct  a gauge
invariant expression for the current operator,
\beq \label{jcov}{\bf j}=\frac{1}{2m}\left(
\psi^\dagger(\frac{1}{i}\bfnabla-e{\bf A})
\psi-\left((\frac{1}{i}\bfnabla+e{\bf A})\psi^\dagger\right)\,\psi\right),\eeq
whose expectation value in terms of the Wigner transforms reads:
\beq
{\bf j}({\bf R}, t)=\int \frac{{\rm d}^3 p}{(2\pi)^3}\,
  \left(\frac{{\bf p}-e{\bf
A}}{m}\right) f({\bf R},{\bf p}, t)=\int\frac{{\rm d}^3 k}{(2\pi)^3}\,
\left(\frac{{\bf k}}{m}\right) \acute f({\bf R},{\bf k}, t).
\eeq
The last expression involving the covariant Wigner function  makes it clear
that ${\bf j}({\bf R}, t)$ is gauge invariant, as it should. The momentum
variable of the gauge covariant Wigner transform is often referred to as the
{\it kinetic}
  momentum. It
is directly related to the velocity of the particles: ${\bf k}=m{\bf v}={\bf
p}-e{\bf A}$.  As for ${\bf p}$, the argument  of the non-covariant Wigner
function, it is related to the gradient operator  and is often 
referred to as the
{\it canonical} momentum.

In order to understand the structure of the equations that we shall obtain for
the QCD plasma, it is  finally instructive to consider the case where the
particles possess internal degrees of freedom (such spin, isospin,
or colour).
The density matrix\index{density matrix} is then a matrix in internal space.
As a specific example, consider a system of spin
$1/2$ fermions. The Wigner distribution reads \cite{BP91}:
\beq
f({\bf p},{\bf R})=f_0({\bf p},{\bf R})+f_a({\bf p},{\bf R})\, \sigma_a,
\eeq
where the $\sigma_a$ are the Pauli matrices,
and the  $f_a$ are three independent
distributions which  describe the  excitations of the system in the
various spin channels; together they form a vector that we can interpret as
a local spin density, ${\bf f}= (1/2) {\rm Tr} (f \bfsigma)$. When
the system is
in a   magnetic field with Hamiltonian
$H=-\mu_0\,\bfsigma\cdot {\bf B}$ the equation of motion for ${\bf f}$ acquires
a new component, $
\del_t {\bf f}= 2\mu_0 {\bf B}\wedge {\bf f},
$
which accounts for the  spin precession in the magnetic field.  In
the linear approximation this precession may be viewed as an extra time
dependence of the distribution function along the characteristics:
\beq\label{spinprecession}
\frac{{\rm d}}{{\rm d}t}=\frac{\del}{\del t}+{\bf v}\cdot\bfnabla_R+2\mu_0 {\bf
B}\wedge\,.
\eeq

It is important to realize that all the differential operators above and in
the Vlasov equation apply to the arguments of distribution 
functions, and not
to the coordinates of the actual particles. Note however that 
equations similar
to the ones presented here  can be obtained for classical  spinning particles.
When the  angular momentum of such particles is large, it can indeed be treated
as a classical degree of freedom, and the corresponding  equations of 
motion have
been written  by Wong
\cite{Wong}. After replacing spin by colour, these equations
have been used by Heinz
\cite{Heinz83,EHPRept} in order to write down transport equations for
classical coloured particles. By implementing the relevant kinematic
approximations  one then recovers \cite{Kelly94}
the non-Abelian Vlasov equations to be derived below,
i.e., (\ref{n0INTRO}) and (\ref{N0INTRO}).
(See also \cite{Gyulassy93,Markov95} for
related work.)

\subsection{QCD Kinetic equations and hard thermal loops }

We are now ready to present the equations that are obtained for the QCD
plasma. These are equations  for generalized one-body
density matrices describing the long wavelength collective motions of colour
particles (quarks and gluons), and possible excitations involving oscillations
of fermionic degrees of freedom. They look formally as the Vlasov
equation\index{Vlasov equation}, the main ones being  \cite{qcd,qed}:
\beq\label{n0INTRO}
\left[ v\cdot D_x,\,\delta n_\pm({{\bf k}},x)\right]=\mp\, g\,{\bf v}
\cdot{\bf E}(x)\,\frac{{\rm d}n_k}{{\rm d}k},\eeq
\beq\label{N0INTRO}
\left[ v\cdot D_x,\,\delta N({{\bf k}},x)\right]=-\, g\,
{\bf v}\cdot{\bf E}(x)\frac{{\rm d}N_k}{{\rm d}k},\eeq
\beq\label{L0INTRO}
(v\cdot D_x){\slashchar\Lambda}({\bf k}, x)=
-igC_f\,(N_k+n_k)\,{\slashchar v}\,\Psi(x).\eeq
In these equations, $v^\mu=(1, {\bf v})$, ${\bf v}={\bf k}/k$, $\Psi(x)$ is an
average (relativistic) fermionic field, and
$\delta n_\pm$, $\delta N$ and ${\slashchar\Lambda}$ are
gauge-covariant Wigner functions for the hard particles. The first two Wigner
functions are those of the density matrices of the quarks and the gluons,
respectively; the last one is  that of a more exotic density matrix which mixes
bosons and fermions degrees of freedom, $\Lambda\sim \langle \psi A\rangle$.
The right hand sides of the equations specify the quantum numbers of the
excitations that they are describing: gluon for the first two, quark 
for the last
one.  One of the major difference between the QCD equations above and the
linear Vlasov equation for QED is the presence of covariant derivatives in the
left hand sides of the equations.  These play a role similar to that of the
magnetic field in (\ref{spinprecession}) for the distribution functions of
particles with spin. (Note that the equation for 
${\slashchar\Lambda}$ holds for
QED, with a covariant derivative there as well.)

The equations (\ref{n0INTRO})--(\ref{L0INTRO}) have a number of interesting 
properties which are
reviewed in \cite{Blaizot:2001nr}. In particular, they
   are covariant under local gauge transformations
of the classical fields, and independent of the gauge-fixing
in the underlying quantum theory.

By solving these equations, one can express the induced sources as
functionals of the background fields. To be specific,
   consider the colour current:
\beq\label{jb1INTRO}
j_a^\mu(x) \equiv 2g\int\frac{{\rm d}^3k}{(2\pi)^3}\,v^\mu
\,{\rm Tr}\,\Bigl(T^a\delta N({\bf k},x)\Bigr),\eeq
where $\delta N$ is the gluon density matrix.
Quite generally, the induced colour current may be expanded in powers
of $A_\mu$,
thus generating the one-particle irreducible amplitudes  of the gauge fields
\cite{qcd}:
\beq\label{exp0}
j^{a}_\mu \,=\,\Pi_{\mu\nu}^{ab}A_b^\nu
+\frac{1}{2}\, \Gamma_{\mu\nu\rho}^{abc} A_b^\nu A_c^\rho+\,...
\eeq
Here, $\Pi_{\mu\nu}^{ab}=\delta^{ab}\Pi_{\mu\nu}$
is the polarization tensor, and the other terms represent
vertex corrections. These  amplitudes are ``hard thermal loops'' (HTL)
\cite{Pisarski89,FT90,Nair91,qed,qcd}\index{hard thermal loop}  which define
the
 effective theory for the soft fields
at the  scale $gT$.
It is worth noticing  that the kinetic equations
isolate  directly these hard thermal loops, in
a gauge invariant manner, without further approximations.

The  gluon density matrix
can be parametrized as in (\ref{DEFW}) :
\beq\label{dn0INTRO}
\delta N_{ab}({\bf k}, x)\,=\,
   - gW_{ab}(x,{\bf v})\,({\rm d}N_k/{\rm d}k),\eeq
where $N_k\equiv 1/({\rm e}^{\beta k}-1)$
is the Bose-Einstein thermal distribution,
and $W(x,{\bf v})\equiv W_a(x,{\bf v}) T^a$
is a colour matrix in the adjoint representation
which depends upon the velocity ${\bf v}={\bf k}/k$
(a unit vector), but not upon the magnitude $k
=|{\bf k}|$ of the  momentum.
Then the colour current takes the form:
\beq\label{jindINTRO}
j_{ind}^\mu(x)=m_D^2\int \frac{{\rm d}\Omega}{4\pi}v^\mu W(x,{\bf v})
\eeq
with $m_D^2\sim g^2 T^2$. A similar representation holds for the
quark density matrices $\delta n_\pm({{\bf k}},x)$.
The kinetic equations for $\delta N_{ab}$ and $\delta n_\pm$ can then 
be  written
as an equation  for
$W_a(x,{\bf v})$:
\bea\label{VLASINTRO}
(v\cdot D_x)^{ab}W_b(x,{\bf v})&=&{\bf v}\cdot{\bf E}^a(x).\eea
They differ  from the corresponding Abelian equation (\ref{EQW})
merely by the replacement of the ordinary   derivative
$\del_x \sim gT$ by the covariant one $D_x=\del_x+igA$.
Accordingly, the soft gluon polarization tensor derived from
 (\ref{jindINTRO})--(\ref{VLASINTRO}),
i.e., the ``hard thermal loop'' $\Pi_{\mu\nu}$,
is formally identical to the photon
polarization tensor obtained from (\ref{EQW}) and
given by (\ref{polarisation2}) \cite{Klimov81,Weldon82a}.
The reason for the existence of an infinite number of hard thermal loops in QCD
is the presence of the covariant derivative in the left hand side of
(\ref{VLASINTRO}). A similar observation can be made by  writing the
induced electromagnetic current in  the form:
\bea
j^\mu_{ind}(x)&=&m_D^2\int \frac{{\rm d}\Omega}{4\pi} v^\mu\int {\rm d}^4y
\,\bra{x}\frac{1}{v\cdot\del}\ket{y}\,{\bf v}\cdot{\bf E}(y)
\nonumber\\&=&\int{\rm d}^4y
\,\sigma^{\mu j}(x,y) E^j(y).
\eea
This expression, which is easily obtained from the expression
(\ref{vlasovlinear}) of $\delta f$, defines the conductivity tensor
$\sigma^{\mu\nu}$. The generalization of this expression to
QCD amounts essentially to
   replacing the ordinary derivative by a covariant one.

\section{Collective phenomena in the quark-gluon plasma}

At the classical level, the effective theory at the scale  $gT$ is summarized
by the generalized Yang-Mills equations
\beq\label{ava}
D_\nu F^{\nu\mu}=\5 m_D^2\int\frac{{\rm d}\Omega}{4\pi}\,
\frac{v^\mu v^i}{v\cdot D}\,E^i
\equiv\hat\Pi_{\mu\nu}^{ab}A_b^\nu
+\frac{1}{2}\, \hat\Gamma_{\mu\nu\rho}^{abc} A_b^\nu A_c^\rho+\,...
\eeq
where the induced current in the right hand side describes
the polarization of the hard particles by the soft colour fields
$A^\mu_a$.
In this equation, $\5 m_D\sim gT$ is the Debye mass\index{Debye mass}, 
$E_a^i$ is the soft electric field, $v^\mu\equiv (1,\,{\bf v})$, and
the angular integral $\int {\rm d}\Omega$ runs over the 
orientations of the unit vector ${\bf v}$. 
The induced current is non-local and  gauge symmetry, which forces the presence
of the covariant derivative $D^\mu=\del^\mu+igA^\mu$
in the denominator of (\ref{ava}), makes it  also non-linear. 

Similarly, the  soft fermionic fields obey the
following generalized Dirac equation \cite{qcd} (with $\5 M\sim gT$ and
$\not\!{v}=\gamma_\mu v^\mu$) :
\beq\label{avpsi}
i\!\not\!\!{D} \psi\,=\,\5 M^2\int\frac{{\rm d}\Omega}{4\pi}
\,\frac {\not\!{v}} {i(v\cdot D)}\,\psi\,
\equiv\, \hat\Sigma\psi + \hat\Gamma_\mu^a A^\mu_a \psi +\,...\eeq

These equations allow  the description of a variety of
collective phenomena. We discuss briefly here some of them ( collective modes,
Debye screening and  Landau damping). More details can be found in  the
lecture by A.~Rebhan. See also \cite{LeBellac96,BIO95}.

\subsection{Collective modes}

The collective plasma waves are propagating solutions to
 (\ref{ava}) or (\ref{avpsi}). We  restrict ourselves in this
subsection to the weak field limit where these equations become linear and
essentially Abelian. 

The solutions can then be analyzed with the help of the
propagator. We  consider here the  gluon propagator\index{gluon propagator}
${}^*\!G_{\mu\nu}$, in Coulomb's gauge  where it has the following non-trivial
components, corresponding
to longitudinal (or electric) and transverse (or magnetic)
degrees of freedom:
\beq\label{DSTAR}
{}^*\!G_{00}(\omega, {\bf p})\equiv {}^*\!\Delta_L(\omega,p),\qquad
{}^*\!G_{ij}(\omega, {\bf p})\equiv
(\delta_{ij}-\hat p_i\hat p_j){}^*\!\Delta_T(\omega,p),\eeq
where:
  \beq\label{effd0}
{}^*\!\Delta_L(\omega,p)\,=\,\frac{- 1}{p^2 + \Pi_L(\omega,p)},\qquad
{}^*\!\Delta_T(\omega,p)\,=\,\frac{-1}{\omega^2-p^2 -\Pi_T(\omega,p)}\,,\eeq
and the electric ($\Pi_L$) and magnetic ($\Pi_T$)
polarization functions are defined as:
\beq\label{PISC}
\Pi_L(\omega, p)\,\equiv\, -\Pi_{00}(\omega, p)\,,\qquad
\Pi_T(\omega,p)\,\equiv\,
\frac{1}{2}\,(\delta^{ij}-\hat p^i \hat p^j)
\Pi_{ij}(\omega,{\bf p})\,.\eeq
Explicit expressions for these functions can be found in \cite{Blaizot:2001nr}.

\begin{figure}
\includegraphics[width=1\textwidth]{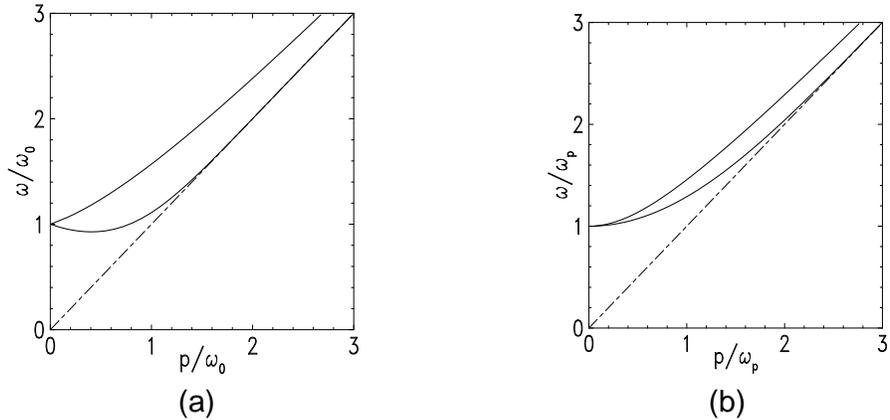}
\caption{Dispersion relation for soft excitations
in the linear regime: (a) soft fermions; 
(b) soft gluons (or linear plasma waves), with
the upper (lower) branch corresponding to transverse (longitudinal)
polarization.}
\label{disp}
\end{figure}


The dispersion relations for the modes are obtained from
the poles of the propagators, that is,
\beq
{p}^2+\Pi_L(\omega_L,p)=0,\qquad \omega_T^2=p^2+\Pi_T(\omega_T,p),
\eeq
for longitudinal and transverse excitations, respectively.
The solutions to these equations,
$\omega_L(p)$ and $\omega_T(p)$,  are displayed in Fig.~\ref{disp}.b.
The longitudinal mode is the analog of the familiar plasma oscillation. It
corresponds to a collective oscillation of the hard particles,
and disappears when $p\gg gT$.
Both dispersion relations are time-like ($\omega_{L,T}(p)>p$),
and show a gap at
zero momentum (the same for transverse and longitudinal modes
since, when $p\to 0$, we recover isotropy).
With increasing momentum, the transverse
branch becomes that of a relativistic particle
with an effective mass $m_{\infty}\equiv m_D/\sqrt{2}$
(commonly referred to as the ``asymptotic mass'').
Although, strictly speaking, the HTL approximation
does not apply at hard momenta, the
above dispersion relation $\omega_T(p)$ remains nevertheless correct
for  $p\sim T$ where it coincides with the light-cone limit
of the full one-loop result \cite{Kraemmer94} :
\beq\label{minfty}
m_\infty^2\,\equiv\,\Pi_T^{1-loop}(\omega^2=p^2)\, =\,
\frac{m_D^2}{2}.\eeq

The dispersion relations of soft fermionic excitations exhibit also collective
feature with a characteristic splitting at low momenta (see
Fig.~\ref{disp}.a). We shall not  discuss here
this interesting phenomenon (see \cite{BIO95} and references therein).

We note finally that particular solutions of the {\it non-linear}
equations (\ref{ava}) have also been found, in
  \cite{prl2,planew,BIO95}. These solutions describe
non-linear plane-waves propagating through the plasma,
and represent truly non-Abelian collective excitations.

\subsection{Debye screening}

The screening of a static chromoelectric field by the plasma
constituents is the natural non-Abelian
generalization of the Debye screening, a familiar phenomenon
in classical plasma physics \cite{PhysKin}.
In coordinate space, screening reduces the range of the gauge
interactions.
In momentum space, it contributes to regulate
the infrared behaviour of the various $n$-point functions.

Screening properties can be inferred from  an analysis of
the effective photon (or gluon) propagators (\ref{effd0})
in the static limit $\omega\to 0$. We have:
\beq\label{static1} \Pi_L(0,p)\,=\,m_D^2\,,\qquad\,\,\,\,
\Pi_T(0,p)=0,\eeq and therefore:
  \beq\label{DSTAT}
{}^*\!\Delta_L(0,p)\,=\,\frac{- 1}{p^2 + m_D^2}\,,\qquad
{}^*\!\Delta_T(0,p)\,=\,\frac{1}{p^2}\,,\eeq
which clearly shows that the Debye mass\index{Debye mass} $m_D$ acts as an
infrared cut-off $\sim gT$ in the electric sector,
while there is no such cut-off in the magnetic sector.

\subsection{Landau damping}

For time-dependent fields, there exists a different screening mechanism
associated to the energy transfer to the plasma constituents.
In Abelian plasmas, this mechanism is known as
{\it Landau damping} \cite{PhysKin}\index{Landau damping}.
The
mechanical work done by a longwavelength electromagnetic field acting
on the charged particles leads to an  energy transfer
  \cite{PhysKin}:
\beq\label{ratew}
\frac{{\rm d}\,{\rm E}_W(t)}{{\rm d}\,t}\,=\,\int {\rm d}^3{\bf x}
\,{\bf E}(t,\,{\bf x})\cdot {\bf j}(t,\,{\bf x}),\eeq
where  $j^i(p)=\Pi_R^{i\nu}(p)A_\nu(p)$ is the induced current.
One can then show that the average energy loss is related to the imaginary part
of the retarded polarization tensor. From (\ref{polarisation2}) we get:
\beq\label{ImPi}
{\rm Im}\, \Pi_R^{\mu\nu}(\omega, {\bf  p})\,=\,-\,
\pi m_D^2\,\omega\int
\frac{{\rm d}\Omega}{4\pi}\,\,v^{\mu}v^{\nu}
\,\delta (\omega - {\bf  v}\cdot{\bf  p})\,.\eeq
The $\delta$-function in (\ref{ImPi}) shows that the particles which
absorb energy are those moving in phase with the field
(i.e., the particles whose velocity component along  ${\bf p}$ is equal
to the field phase velocity: ${\bf v}\cdot \hat{\bf p} =\omega/p$).
Since in ultrarelativistic plasmas ${\bf v}$ is a unit vector,
  only {\it space-like} ($|\omega| < p$) fields
are damped in this way.

To see how this mechanism leads to screening,
consider the effective photon (or gluon) propagator  (\ref{effd0}), and focus
on the magnetic propagator. For small but non-vanishing frequencies the
corresponding polarization function  $\Pi_T(\omega,p)$ is dominated by its
imaginary part:
\beq\label{IMPIT}
\Pi_T(\omega \ll p)\,=\,-i\,\frac{\pi}{4}\,m_D^2\,\frac{\omega}{p}
+\,{\rm O}(\omega^2/p^2)\,,
\eeq
and therefore
\beq\label{Vt}
{}^*\!\Delta_T(\omega\ll p)\simeq\,\frac{1}
{p^2-i\,(\pi \omega/4p)\,m_D^2}.\eeq
Thus Im$\,\Pi_T(p)$ acts as
a frequency-dependent IR cutoff at momenta
$p\sim (\omega m_D^2)^{1/3}$. That is, as long as the frequency
$\omega$ is different from zero, the soft momenta are
dynamically screened by Landau damping \cite{Baym90}.

\section{The entropy of the quark-gluon plasma}

We come now to the last part of these lectures which will be mainly 
devoted to an
introduction to the recent progress made in the calculation of the 
entropy of the
quark-gluon plasma. We first comment on various aspects of perturbation theory
and show that it is not appropriate for calculating the thermodynamics of the
quark-gluon plasma, even a high temperature where the coupling is weak. The
main source of difficulties  is that the contributions of the collective modes,
for which we have constructed an effective theory in the previous sections,
are non perturbative and cannot be 
expanded  in powers of the coupling constant.  We then show that these
contributions can be included by using self-consistent approximations familiar
in many-body physics. These are best formulated for the entropy of the plasma,
for which we obtain a simple approximation which provides an accurate
description of lattice gauge calculations.   

\subsection{Results from perturbation theory}

  The free energy has been calculated up to order $g^5$, including the
contribution of fermions \cite{Arnold94}. However, since our purpose here is
mostly  pedagogical, we
shall limit our discussion to  the gluon contribution at
order $g^4$, in an SU(N) gauge theory. The pressure
$P=-F/V$ can then be written:
\beq\label{pressure_QCD}
P=P_0\left[1+a_2 g^2+a_3 g^3+ \left(a_4(\mu/T)+a'_4\ln g\right) g^4 
+O(g^5)\right],
\eeq
with
\bea\label{coef_pressure}
\lefteqn{
a_2=-5 \left( \frac{\sqrt{N}}{4\pi }\right) ^{2}, \,
a_3= \frac{80}{\sqrt{3}}\left( \frac{\sqrt{N}}{4\pi }\right) ^{3},\,
a_4'= 240\left( \frac{\sqrt{N}}{4\pi }\right) ^{4}\ln
\frac{\sqrt{N}}{2\pi
\sqrt{3}}  }\nonumber \\
  & & a_4= -5\left( \frac{\sqrt{N}}{4\pi }\right) ^{4}\left[ 
\frac{22}{3}\ln \frac{
{\mu }}{4\pi T}+\frac{38}{3}\frac{\zeta ^{\prime }(-3)}{\zeta (-3)}-
\frac{148}{3}\frac{\zeta ^{\prime }(-1)}{\zeta (-1)}-4\gamma _{E}+\frac{64}{5}
\right],\nonumber\\
& &
\eea
 where $\zeta$ is Riemann's zeta function, and $\mu$ the renormalization scale.

The first term in the expansion is $P_0$, the pressure of an ideal 
gas of gluons:
\beq
P_0=(N^2-1)\;T^{4}\frac{\pi ^{2}}{45}.
\eeq
  The next term, of order $g^{2}$, is a genuine perturbative 
correction, and so is the term of order $g^4$.
  The  contributions of order $g^{3}$  can be interpreted as
a contribution of the collective modes to the
pressure, and the odd power reflects the fact that the
calculation of this  contribution requires resummations. Similar 
resummations are
responsible for the term in $g^4 \ln g$.

We note that some of the coefficients in (\ref{coef_pressure}) depend
  on the renormalization scale
$\mu$. However,  the pressure  itself should  not depend on $\mu .$ It
obeys a renormalization group equation\index{renormalization group}:
\beq
\left[ \mu^2 \frac{\partial }{\partial \mu^2 }+\left( \mu^2 
\frac{{\rm d}\alpha}{%
{\rm d} \mu^2}\right) \frac{\partial }{\del \alpha}\right]P=0.
\eeq
In this equation,  $\alpha(\mu) \equiv g^2(\mu)/4\pi$ is  the
running  coupling constant which satisfies the equation:
\beq\label{beta_function}
\mu^2\frac{{\rm d}\alpha}{{\rm d}\mu^2}=\beta(\alpha)=-\beta_2 
\alpha^2-\beta_3 \alpha^3,
\eeq
with
\beq\label{beta2}
\beta _{2} =\frac{11N}{12\pi },\qquad\qquad \beta_3= \frac{17 N^2}{24\pi^2} .
\eeq
One can then show that, indeed, $P$ is   independent of ${\mu
}$: the explicit $\mu$ dependence of the coefficients cancels with 
that of the running coupling. Look
indeed at the following combination of terms coming from the 
contributions of $a_2 g^2$ and  the $\mu$
dependent part of $a_4 g^4$:
\beq
\frac{N}{4\pi }\left\{ \alpha +\frac{N}{4\pi }\alpha ^{2}\frac{22}{3}%
\ln \frac{{\mu }}{4\pi T}\right\} .
\eeq
By taking the derivative of this expression with respect to $\mu^2 $
one gets:
\beq
\mu ^{2}\frac{\rm d}{{\rm d} \mu ^{2}}\left\{ {}\right\} =\mu ^{2}\frac{
{\rm d}\alpha }{{\rm d} \mu ^{2}}+\frac{N}{4\pi }\alpha ^{2}\frac{11}{%
3}+{\rm \ higher\;order\;terms.}
\eeq
By using the leading order expression of the $\beta$-function given 
in (\ref{beta_function}), one
then obtains, as announced:
\beq
-\frac{11}{12\pi }N\;\alpha ^{2}+\frac{N}{4\pi }\alpha ^{2}\frac{11}{%
3}=0.
\eeq

Note however, that the pressure is only
   {\it formally} independent of $\mu $ at order $g^{4}$, in the sense 
that its derivative
with respect to $\mu$ involves terms of order $g^5$ at least. But the 
approximate expression
(\ref{pressure_QCD}) for
$P$ does depend on
$\mu$. As in all perturbative calculations, one is then led to look 
for the best value of $\mu$, i.e. the
one which  minimizes the  higher order corrections. In the present 
context, a ``natural choice'' is to fix
  $\mu=2\pi T$, where $2\pi T$ is the  scale  provided by the
basic Matsubara frequency. This choice makes the running coupling 
decrease with increasing temperature, and
leads in particular to the expectation that the quark-gluon plasma 
becomes perturbative at very large
temperature.

By calculating explicitly the various coefficients in (\ref{coef_pressure})
for $N=3$, one can write (\ref{pressure_QCD})  thus:
\bea
\lefteqn{ P=P_0\left[ 1- 0.095 g^2 +0.12 g^3 + \right.}\nonumber\\ & 
&  \left.\left(0.09 \ln
g-0.007-0.013\ln\left(\frac{\mu}{2\pi T}\right)\right) g^4 +O(g^5) \right].
\eea
Then, if for example one fixes $\mu=2\pi T$ and chooses a large temperature
such  that
$\alpha(2\pi T)=0.1$, one gets
$g=1.12$, and
\bea
P=P_0\left[ 1-0.12+0.17+0.004\right],
\eea
which shows no sign of convergence, with the term of order $g^3$ larger than
the term of order $g^2$. Furthermore, if one analyzes the dependence of $P$ on
the renormalization scale, on finds large variations as $\mu$ runs within the
interval 
  $\pi T <\mu< 4\pi T$.

Attempts have been made to extract information from the first terms of this
series using Pade approximants \cite{Hatsuda:1997wf,Chiku:1998kd}  or Borel
summation techniques \cite{Parwani:2001rr,Parwani:2001am}. The resulting
expression of the pressure becomes indeed a smooth function of the coupling,
better behaved than the polynomial approximation (\ref{pressure_QCD}). These
techniques however, which are in some situations very powerful, provide little
physical insight, and we shall not discuss them further here.

The behavior of perturbation theory does not improve as one takes into
account the higher order terms that one can calculate (namely orders $g^4$
and $g^5$). Furthermore, at order $g^6$, as we have already mentioned,
perturbation theory becomes inapplicable because of infrared divergences. It
has been shown in 
\cite{Braaten:1995na,Braaten:1996ju,Braaten:1996jr} how, in principle, an
effective theory\index{effective theory} could be constructed to overcome this
particular problem by marrying analytical techniques (to determine the
coefficients of the effective theory) and numerical ones (to solve the non
perturbative 3-dimensional effective theory). The resulting effective theory
is a 3-dimensional theory of static fields, with Lagrangian:
\beq
{\cal L}_{eff}=\frac{1}{4}
(F^a_{ij})^2+\frac{1}{2}(D_iA_0^a)^2+\frac{1}{2}m_D^2(A_0^a)^2+\lambda 
(A_0^a)^4
+\delta{\cal L},
\eeq
with $
D_i=\partial_i-ig\sqrt{T}A_i
$. This strategy has been applied recently to the calculation of the free
energy of the quark-gluon plasma a high temperature \cite{Kajantie:2001iz}.
The slow convergence of the pressure towards the ideal gas value, that is seen
in lattice calculations above $T_c$, is well reproduced. It is
worth-emphasizing that this technique of dimensional
reduction\index{dimensional reduction} puts a special weight on the static
sector (it singles out the contributions of the zero Matsubara frequency).
However, as we shall see, it may be advantageous to keep, even in the
calculation of equilibrium thermodynamic properties, the full spectral
information that one has about the plasma excitations. 

There are indeed indications that lattice data are well accounted for by
simple phenomenological models of weakly interacting quasiparticles
\cite{Peshier,LH}.  
In the case of the scalar field, the dominant effect of the interactions is
to give a mass to the excitations. An indeed   a perturbative expansion in
terms of screened propagators (that is keeping the screening mass $\sim gT$ as
a parameter, not  considered as a term of order $g$ entering the expansion) 
has been shown to be quite stable  with good
convergence properties
\cite{Karsch97}. In the case of gauge theory, the effect of the interactions
is more complicated than just generating a mass. But we know how to determine
the dominant corrections to the self-energies. When the momenta are soft, these
are given by the hard thermal loops discussed above. By adding these
corrections to the tree level Lagrangian, and subtracting them from the
interaction part, one generated the so-called hard thermal loop perturbation
theory \cite{ABS}. The resulting perturbative expansion is made complicated
however by the non local nature of the hard thermal loop action, and by the
necessity of introducing temperature dependent counter terms. At the expense
of some extra formalism, some of these difficulties can be avoided. This is
discussed now.

\subsection{Skeleton expansion for thermodynamic potential and entropy}

In this section we recall the formalism of propagator renormalization   that
allow systematic rearrangements of the perturbative expansion while avoiding
double-counting.  We shall see in particular how self-consistent
approximations can be used to obtain a simple expression for the entropy which
isolates the contribution of the elementary excitations as a leading
contribution. For pedagogical
purposes, we  shall mainly consider in these lectures the example of the
scalar field.

The thermodynamic potential $\O=-PV$ of the  scalar field can be written as the
following  functional of the full propagator $D$ \cite{LW,Baym}:
\beq
\label{LW}
\b \O[D]=-\log Z=\2 \Tr \log D^{-1}
-\2 \Tr \Pi D+\Phi[D]\,,
\eeq
where $\Tr$ denotes the trace in configuration space,
$\b=1/T$, $\Pi$ is the self-energy related to $D$ by
Dyson's equation ($D_0$ denotes the bare propagator):
\beq\label{Dyson}
D^{-1}=D^{-1}_0+\Pi,
\eeq
and $\Phi[D]$ is the sum of the 2-particle-irreducible ``skeleton''
diagrams
\beq\label{skeleton}
-\Phi[D]=
\includegraphics[bb = 50 390 550 430,width=7truecm]{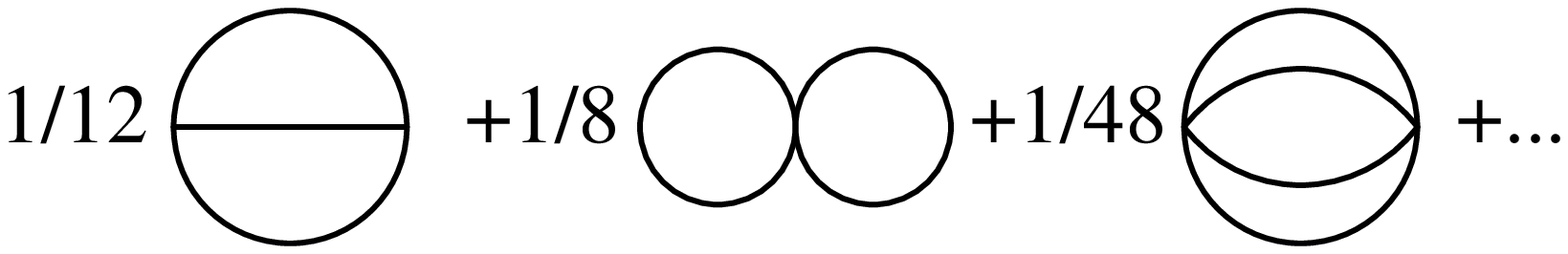}
\eeq

\vspace*{0.1cm}
The essential property of the functional $\O[D]$ is to be
stationary under variations of $D$ (at fixed $D_0$) around the
physical propagator. The physical pressure is then obtained
as the value of $\O[D]$ at its extremum. The stationarity condition,
\beq\label{selfcons}
{\d\O[D] / \d D}=0,
\eeq
implies the following relation
\beq\label{PhiPi}
\d\Phi[D]/\d D=\2\Pi,
\eeq
which, together with (\ref{Dyson}), defines the
physical propagator and self-energy in a self-consistent way.
The equation (\ref{PhiPi}) expresses the fact that the skeleton
diagrams contributing to $\Pi$ are
obtained by opening up one line of a two-particle-irreducible skeleton.
Note that while the diagrams of the bare perturbation theory,
i.e., those involving
bare propagators, are counted once and only once in the expression of
$\Pi$ given above, the
diagrams of bare perturbation theory
contributing to the thermodynamic potential
are counted several times in
$\Phi$. The extra terms in (\ref{LW}) precisely correct for
this double-counting.

Self-consistent (or variational)
approximations\index{self-consistent approximation}, i.e., approximations which
preserve the stationarity property (\ref{selfcons}),
are  obtained by selecting a class of skeletons in
$\Phi[D]$ and calculating $\Pi$ from (\ref{PhiPi}).
Such approximations are commonly called ``$\Phi$-derivable''
\cite{Baym}.

The traces over configuration space in (\ref{LW}) involve integration over
imaginary time and over spatial coordinates. Alternatively, these can be
turned into summations over Matsubara frequencies and integrations over
spatial momenta:
\beq
\int_0^\beta {\rm d}\tau
\int {\rm d}^3x \rightarrow \beta V
  \int [{\rm d}k] ,\eeq
where $V$ is the spatial volume,
  $k^\mu=(i\omega_n, {\bf k})$
and $\omega_n = n\pi T$, with $n$ even (odd) for bosonic
(fermionic) fields (the fermions will be discussed later).
We have introduced a condensed notation for the
the measure of the loop integrals (i.e., the sum over the
Matsubara frequencies $\omega_n$ and the integral over the
spatial momentum ${\bf k}$):
\beq\label{mesure}
\int[{\rm d}k]\equiv T\sum_{n, even} \int\frac{{\rm d}^3k}{(2\pi)^3}
\eeq
Strictly speaking, the sum-integrals in equations like  (\ref{LW})
contain  ultraviolet divergences, which requires regularization
(e.g., by dimensional continuation). Since, however,
most of the forthcoming calculations will be free
of ultraviolet problems, we do not need
to specify here the UV regulator  (see however Sect.~\ref{secsimplemodel} for
explicit calculations).

For the purpose of developing approximations for  the entropy
it is convenient to
perform the  summations over the Matsubara frequencies. One obtains  then
integrals over real frequencies involving discontinuities of propagators or
self-energies which have a direct physical significance. Using standard
contour integration techniques, one gets:
\bea\label{Omega(D)}
\Omega/V=\int\!\!{d^4k\0(2\pi)^4}\, n(\omega)\left(\Im \log(-\omega^2+k^2+\Pi)
-\Im\Pi D\right)+T\Phi[D]/V\nonumber\\& &
\eea
where $n(\omega)=1/({\rm e}^{\beta \omega}-1)$.

  The analytic propagator $D(\omega,k)$ can be
expressed in terms of the spectral function\index{spectral function}:
\beq
D(\o,k)=\int_{-\infty}^\infty {dk_0\02\pi}{\rho(k_0,k)\0k_0-\o}.
\eeq
and we define, for $\omega$ real,
\beq
\Im D(\omega,k)\equiv \Im D(\omega+i\epsilon,k)=\frac{\rho(\omega,k)}{2}.
\eeq
  The imaginary parts of other quantities are defined
similarly.

We are now in the position to calculate  the entropy\index{entropy} density:
\beq
{\cal S}=-{\6(\O/V)/\6T}\,.
\eeq
The thermodynamic potential, as given by (\ref{Omega(D)}) depends on the
temperature through the statistical factors $n(\omega)$ and the spectral
function $\rho$, which is determined entirely by the self-energy.
Because of  (\ref{selfcons}) the
temperature derivative of the spectral
density in the dressed propagator cancels
out in the entropy density and one obtains
\cite{Riedel,VB}:
\bea\label{Ssc}
{\cal S}&=&-\int\!\!{d^4k\0(2\pi)^4}{\6n(\o)\0\6T} \Im \log D^{-1}(\o,k) \nn
&&+\int\!\!{d^4k\0(2\pi)^4}{\6n(\o)\0\6T} \Im\Pi(\o,k) \Re D(\o,k)+{\cal S}'
\eea
with
\beq\label{SP0}
{\cal S}'\equiv -{\6(T\Phi)\0\6T}\Big|_D+
\int\!\!{d^4k\0(2\pi)^4}{\6n(\o)\0\6T} \Re\Pi\, \Im D.
\eeq
For the two-loop skeletons, we have:
\beq\label{Sprime}
{\cal S}'=0.
\eeq
Loosely speaking, the first two terms in (\ref{Ssc})
represent essentially the entropy of ``independent quasiparticles'',
while ${\cal S}'$
accounts for a residual interaction among these quasiparticles
\cite{VB}.

The vanishing of ${\cal S}'$  holds
whether the propagator are the self-consistent propagators or not.
That is,
only the relation (\ref{PhiPi}) is used in the proof which does not require $D$
to satisfy the self-consistent Dyson equation (\ref{Dyson}).
A general analysis of the contributions to
${\cal S}'$ and their physical
interpretation can be found in \cite{CP2}.

We emphasize  now a few attractive features of the formula (\ref{Ssc})
with ${\cal S}'=0$,
which makes the entropy a privileged quantity to study the thermodynamics of
ultrarelativistic plasmas. We note first that  the formula for
$\cal S$ at 2-loop order
involves the self-energy only at 1-loop order.  Besides this important
simplification, this formula for
${\cal S}$, in contrast to the pressure, has the
advantage of manifest ultra-violet finiteness, since ${\6n/\6T}$
vanishes exponentially for both $\o\to\pm\infty$.
Also, any multiplicative renormalization $D\to ZD$, $\Pi\to Z^{-1}\Pi$
with real $Z$ drops out from (\ref{Ssc}). Finally, the entropy has  a more
direct quasiparticle interpretation than the pressure.
This will be illustrated
explicitly in the simple model of the next subsection.

\subsection{A simple model}
\label{secsimplemodel}

In this section we shall present the  self-consistent solution
for the $(\lambda/4!) \phi^4$ theory, keeping in $\Phi$ only the two-loop
skeleton.
Anticipating the fact that the fully dressed propagator will be that
of a massive particle, we write the spectral function as
$\rho(k_0,{\bf k})=2\pi \,\epsilon(k_0)\,\delta(k_0^2-{\bf k}^2-m^2)$,
and consider $m$ as a variational parameter. The thermodynamic potential
(\ref{LW}), or equivalently the pressure, becomes then a simple
function of $m$.
By Dyson's equation, the self-energy is simply $\Pi=m^2$. We set:
\beq\label{IM}
I(m)\,\equiv \,\frac{1}{2}\int [{\rm d}k]\, D(k)\,= \2\int [{\rm d}k]
\,\frac{1}{\omega_n^2 + {\bf k}^2 + m^2}.
\eeq
Then  the pressure $P=-\Omega/V$
can be written as:
\bea\label{Pscal}
-P=\frac{1}{2}\int\frac{{\rm
d}^3k}{(2\pi)^3}\varepsilon_k+\frac{1}{\beta}\int\frac{{\rm
d}^3k}{(2\pi)^3}\log(1-{\rm e}^{-\beta\varepsilon_k})
-m^2 I(m)+\frac{\lambda_0}{2} I^2(m),\nonumber\\&&
\eea
where
   $\varepsilon_k^2\equiv k^2+m^2$.
By demanding that $P$ be stationary with respect to $m$
one obtains the self-consistency condition which takes
here the form of a ``gap equation'':
\beq
\label{gapp}
m^2\,=\,{\lambda_0}\,I(m).
\eeq
{The pressure in the two-loop $\Phi$-derivable approximation,
as given by (\ref{IM})--(\ref{gapp}), is formally the same
as the pressure per scalar degree of freedom in the (massless)
$N$-component model with the interaction term written as
${3\0N+2}(\lambda/4!)(\phi_i\phi_i)^2$ in the limit $N\to \infty$
\cite{Drummond:1998cw}. From the experience with this latter model, we know
that (\ref{IM})--(\ref{gapp}) admit an exact, renormalizable solution
which we recall now.
}

At this stage, we need to specify some properties of the loop integral $I(m)$
which  we can write as  the sum of a vacuum piece $I_0(m)$ and a finite
temperature piece $I_T(m)$ such that, at fixed $m$,
$I_T(m)\to 0$ as $T\to 0$.
We use dimensional regularization
to control the ultraviolet divergences present in $I_0$,
which implies $I_0(0)=0$. Explicitly
one has:
\beq\label{I(m)regul}
\mu^{\epsilon}
I(m)=-\frac{m^2}{32\pi^2}
\left(\frac{2}{\epsilon}+\log\frac{\bar\mu^2}{m^2}+1\right)
+I_T(m)+{\rm O}(\epsilon),
\eeq
with
\beq\label{ITm}
I_T(m)=\int\!\!{d^3k\0(2\pi)^3}\,\frac{n(\varepsilon_k)}
{2\varepsilon_k}\,,
\eeq
and
$\varepsilon_k\equiv (k^2+m^2)^{1/2}$. In (\ref{I(m)regul}), $\mu$ is
the scale of dimensional regularization, introduced, as usual,
by rewriting the bare coupling $\lambda_0$
as $\mu^{\epsilon}\hat\lambda_0$, with dimensionless
$\hat\lambda_0$; furthermore, $\epsilon=4-n$, with $n$ the number of
space-time dimensions, and
$\bar\mu^2=4\pi{\rm e}^{-\gamma}\mu^2$.

We use the modified minimal subtraction
scheme ($\overline{\hbox{MS}}$)
and define a dimensionless renormalized coupling
$\lambda$ by:
\beq\label{RENL}
\frac{1}{\lambda}=\frac{1}{\lambda_0\mu^{-\epsilon}}+\frac{1}{16\pi^2\epsilon}.
\eeq
When expressed in terms of the renormalized coupling, the gap
equation becomes free of ultraviolet divergences. It reads:
\beq\label{GAP2}
m^2\,=\,\frac{\lambda}{2}\int\frac{{\rm d}^3k}{(2\pi)^3}\,
\frac{n(\varepsilon_k)}{\varepsilon_k}\,+\,
\frac{\lambda  m^2}{32 \pi^2}\left(\log \frac{m^2}{\bar\mu^2}
\,-1\right),
\eeq
The renormalized coupling constant satisfies
\beq\label{RGSC}
\frac{{\rm d}\lambda}{{\rm d}\log \bar\mu}\,=\,\frac{\lambda^2}{16\pi^2},
\eeq
which ensures that the solution $m^2$ of (\ref{GAP2})
is independent of $\bar\mu$. The expression (\ref{RGSC}) coincides with the
exact
$\beta$-function in the large-$N$ limit,
but gives only one third of the lowest-order perturbative
$\beta$-function for $N=1$. 
This is no actual fault since the running
of the coupling affects the thermodynamic potential only at
order $\lambda^2$ which is beyond the perturbative accuracy of
the 2-loop $\Phi$-derivable approximation. In order to see the
correct one-loop $\beta$-function at finite $N$, the approximation
for $\Phi$ would have to be pushed to 3-loop order. 

Note also that, in the present approximation, the
renormalization (\ref{RENL}) of the coupling constant
is sufficient to make the pressure (\ref{Pscal}) finite.
Indeed, in dimensional regularization
the sum of the zero point energies ${\varepsilon_k}/2$
in (\ref{Pscal}) reads:
\beq\label{zeropoint}
\mu^{\epsilon}\int\frac{{\rm d}^{n-1}k}{(2\pi)^{n-1}}\,\,
\frac{\varepsilon_k}{2}\,=\,-\frac{m^4}{64\pi^2}
\left(\frac{2}{\epsilon}+\log\frac{\bar\mu^2}{m^2}+{3\02}\right)
+{\rm O}(\epsilon),
\eeq
so that
\beq
\mu^\epsilon\int\frac{{\rm d}^{n-1}k}{(2\pi)^{n-1}}\,\,
\frac{\varepsilon_k}{2}\,-\,
\frac{\Pi^2}{2\hat\lambda_0}\,=\,-\,{m^4\02\lambda}\,-\,
\frac{m^4}{64\pi^2}
\left(\log\frac{\bar\mu^2}{m^2}+{3\02}\right)+{\rm O}(\epsilon)\eeq
is indeed UV finite as $n\to 4$. {After also using the
gap equation (\ref{GAP2}), one obtains the $\bar\mu$-independent
result}
\beq\label{PPHI}
P=-T\int\frac{{\rm
d}^3k}{(2\pi)^3}\,\log(1-{\rm e}^{-\beta\varepsilon_k})
+{m^2\02}I_T(m)+{m^4\0128\pi^2}.
\eeq

We now compute the entropy according to (\ref{Ssc}). Since
  $\Im\Pi=0$ and $\Re\Pi=m^2$, we have simply:
\beq\label{Sscphi4}
{\cal S}=-\int\!\!{d^4k\0(2\pi)^4}{\6n(\o)\0\6T}\Im\log (k^2-\o^2+m^2).
\eeq
Using
\beq\label{IM0}
  \Im\log(k^2-\o^2+m^2) \,=\,-\pi\epsilon(\o)\theta(\o^2-
\varepsilon_k^2),
\eeq
and the identity,
\beq\label{ID1}
  \frac{\del n(\o)}{\del T}=-\,{\6\sigma(\o)\0\6\o},\qquad\sigma(\o)
\equiv -n\log n + (1+n)\log(1+n), \eeq
   one can rewrite (\ref{Sscphi4})  in the  form
(with $n_k\equiv
n(\varepsilon_k)$):
\beq\label{SPHI}
{\cal S} =\int\!\!{d^3k\0(2\pi)^3}\Bigl\{(1+n_k)\log (1+n_k) \,-\,
n_k\log n_k\Bigr\}.\eeq
This formula   shows that,
in the present approximation, the entropy of the interacting scalar
gas is formally identical to the entropy of an ideal gas of massive
bosons, with mass $m$.

It is instructive to observe that such a simple
interpretation does not hold for the pressure. The pressure of
an ideal gas of massive bosons is given by:
\bea\label{Pmassbosons}
P^{(0)}(m)&=&\int\!\!{d^3k\0(2\pi)^3}\int_{\epsilon_k}^\infty{\rm
d}\o\left(n(\o)+\frac{1}{2}\right)\nonumber\\
&=&-\int\!\!{d^3k\0(2\pi)^3}\left\{T\log(1-{\rm
e}^{-\epsilon_k/T})+\frac{\epsilon_k}{2}\right\},
\eea
which differs indeed from (\ref{Pscal}) by the term $m^4/\lambda$ which
corrects for the double-counting of the interactions included in the
thermal mass. 

\subsection{Comparison with thermal perturbation theory}
\label{secscapprox}

In view of the subsequent application to QCD, where a  fully
self-consistent determination of the gluonic self-energy  seems
prohibitively difficult, {we shall be led to consider
approximations to the gap equation. 
These 
will be constructed such that they reproduce
(but eventually transcend)
the perturbative results up to and including
order $\lambda^{3/2}$ or $g^3$, which
is the maximum perturbative accuracy allowed
by the approximation  ${\cal S}'=0$.

In view of this it is important to understand the perturbative
content of the
self-consistent approximations
for $m^2$, $P$ and ${\cal S}$.
In this section we shall demonstrate that, when expanded in powers
of the coupling constant, these approximations reproduce
the correct perturbative results up to order $\lambda^{3/2}$
\cite{Kapusta}. This will also elucidate how perturbation
theory gets reorganized by the use of the skeleton representation
together with the stationarity principle.

For the scalar theory with only $(\lambda/4!)\,\phi^4$ self-interactions,
we write\footnote{This normalization for $g$ is chosen in
view of the subsequent extension to QCD since it makes
the scalar thermal mass in (\ref{SCMHTL}) equal to
the leading-order Debye mass in pure-glue QCD.}
$\lambda\equiv 24g^2$, and compute the corresponding
self-energy $\Pi= m^2$ by solving the gap equation (\ref{GAP2})
in an expansion in powers of $g$, up to order $g^3$.
Since we anticipate $m$ to be of order $gT$, we can ignore
the second term $\propto \lambda m^2 \sim g^4$ in the r.h.s.
of (\ref{GAP2}), and perform a high-temperature expansion of
the integral $I_T(m)$
in the first term (cf. (\ref{ITm})) up to terms linear in $m$.
This gives the following, approximate, gap equation:
\beq\label{APPROXGAP}
  m^2\,\simeq\,g^2T^2-{3\0\pi}\,g^2Tm\,.\eeq
The first term in the r.h.s.\ arises as
\beq
\label{SCMHTL}
24 g^2I_T(0)\,=\,12g^2
\int\!\!{d^3k\0(2\pi)^3}\,\frac{n(k)}{k}\,=\,g^2T^2\,\equiv\,\hat m^2.
\eeq
This is also the leading-order result for $m^2$,
commonly dubbed
the ``hard thermal loop''.

The second term, linear in $m$, in (\ref{APPROXGAP}) comes from
\bea
&&12 g^2 \int\!\!{d^3k\0(2\pi)^3}\left(
\frac{n(\varepsilon_k)}{\varepsilon_k}-\frac{n(k)}{k}\right)
\,\simeq \,\nn\\&&\qquad\qquad
12 g^2 T\int\!\!{d^3k\0(2\pi)^3}\left(\frac{1}
{k^2+ m^2}-\frac{1}{k^2}\right)\,=\,
-\frac{3g^2}{\pi} mT\,,
\eea
where we have used the fact that
the momentum integral is saturated by soft
momenta $k\sim gT$, so that to the order of interest
$n(\varepsilon_k)\simeq T/\varepsilon_k$ (and similarly
$n(k)\simeq T/k$).
This provides the next-to-leading order (NLO)
correction to the thermal mass
\beq\label{SCNLO}
\delta m^2\,\equiv\,-\frac{3g^2}{\pi}\hat mT\,=\,-\frac{3}{\pi}\,g^3T^2\,.
\eeq

Thus, to order $g^3$, one has $m^2=\hat m^2+\delta m^2$.
In standard perturbation theory\cite{Kapusta,LeBellac96},
the first term arises as the
one-loop tadpole diagram evaluated with a bare massless propagator,
while the second term comes from the same diagram where
the internal line is soft and dressed by the HTL, that is
$\hat D(\o,k) \equiv -1/(\o^2 - k^2 -\hat m^2)$.

Consider similarly the perturbative estimates for the pressure and
entropy, as obtained
by evaluating (\ref{Pscal}) and (\ref{SPHI})
with the perturbative self-energy $\Pi =m^2
\simeq \hat m^2+\delta m^2$, and further
expanding in powers of $g$, to order $g^3$.
The renormalized version of
(\ref{Pscal}) yields, to this order
(recall that $m\sim gT$ and $\lambda\sim g^2$),}
\beq\label{Pdem}
P\,\simeq\,
\frac{\pi^2 T^4}{90}-\frac{m^2 T^2}{24}+\frac{m^3T}{12\pi}+\cdots
+\frac{m^4}{2\lambda}.
\eeq
The first terms before the dots represent the pressure of massive bosons,
i.e.  (\ref{Pmassbosons}) expanded up to third order in powers of
$m/T$. From  (\ref{Pdem}), it can be easily verified that the
above perturbative solution for $m^2$  ensures the stationarity of
$P$ up to order $g^3$, as it should. Indeed, if we denote
\beq
  P_2(m) \,\equiv\,-\frac{m^2 T^2}{24}\,+\,\frac{m^4}{2\lambda}\,,
\qquad\, P_3(m)\,\equiv\, \frac{m^3 T}{12\pi}\,,\eeq
then the following identities hold:
\beq\label{STAT2}
{\del P_2\0\del m}\bigg|_{\5m}\,=\,0,\qquad
{\del P_2\0\del m}\bigg|_{\5m+\delta m}\,+\,
{\del P_3\0\del m}\bigg|_{\5m}\,=\,0.\eeq
This shows that the NLO mass correction $\delta m^2\sim g^3T^2$
can be also obtained as
\beq
\delta m^2\,=\,-\,{(\del P_3/\del m)\0(\del^2 P_2/\del m^2)}\bigg|_{\5m}
\,=\,-\frac{3g}{\pi}\,\5m^2\,,\eeq
in agreement with  (\ref{SCNLO}).
Moreover, $P_2\equiv  P_2(\5m) = -g^2T^2/48$ and $P_3\equiv  P_3(\5m) =
\5m^3 T/12\pi$ are indeed the correct perturbative corrections
to the pressure, to orders $g^2$ and $g^3$, respectively
\cite{Kapusta}. In fact, the pressure to this order can be
written as:
\bea\label{pressurescal}
P&=&\frac{\pi^2 T^4}{90}-\frac{\hat m^2
T^2}{24}\,(1-\frac{3}{\pi}\,g)+\frac{\hat m^3T}{12\pi}+\cdots
\,+\,\frac{\hat m^4}{2\lambda}\,(1-\frac{3}{\pi}\,g)^2 +{\cal 
O}(g^4)\nonumber\\
&=&\frac{\pi^2 T^4}{90}-\frac{\hat m^2}{48} T^2+\frac{\hat m^3 T}{12\pi}\,.
\eea
Note that the term of order $g^2$ is only {\it half}
of that one would obtain from (\ref{Pmassbosons})
by replacing $m$ by $\hat m$. This is due to the 
mismatch between (\ref{Pmassbosons}) and the correct expression (\ref{Pscal})
for the pressure.
In fact the net
order $g^2$ contribution to the pressure comes from $\Phi$ evaluated with
bare propagators: the order $g^2$ contributions in the other two terms
mutually cancel indeed. This is to
be expected: there is a single diagram of order $g^2$; this is a skeleton
diagram, counted therefore once and only once in $\Phi$.
Observe also that the
terms of order $g^3$ originating from  the terms  $\hat m^2$ and
$\hat m^4$ mutually cancel; that is, the NLO mass correction
$\delta m$ drops out from the pressure up to order
$g^3$. This is no accident: the cancellation
results from the stationarity of $P$ at order $g^2$, the first equation
(\ref{STAT2}).

Consider now the entropy density. The correct perturbative result up to order
$g^3$ may be obtained directly by taking the total derivative of the
pressure, (\ref{pressurescal}) with respect to
$T$. One then obtains:
\beq\label{SPERTSC}
{\cal S}=\frac{4}{T}\left( \frac{\pi^2 T^4}{90}
-\frac{\hat m^2 T^2}{48}+\frac{\hat m^3 T}{12\pi}\right)+{\cal O}(g^4).
\eeq

We wish, however, to proceed differently, using (\ref{SPHI}), or
equivalently,
  since $\del P/\del m=0$ when $m$ is a solution of the
gap equation, by writing:
\beq
{\cal S}=\left.\frac{\del P}{\del T}\right|_m.
\eeq
This yields:
\beq\label{Sdem}
{\cal S}=\frac{4}{T}\left( \frac{\pi^2 T^4}{90}-\frac{m^2 T^2}{48}+\frac{m^3
T}{48\pi}\right)+{\cal O}(m^4/T),
\eeq
which coincides as expected with the expression obtained
by expanding the entropy (\ref{SPHI}) of massive bosons,
up to order $(m/T)^3$.
If we now
replace $m$ by its leading order value
$\hat m$, the resulting approximation 
for ${\cal S}$
reproduces the perturbative effect of order $\sim g^2$,
but it underestimates the correction of order $g^3$ by a factor of 4.
This is corrected  by changing $m$ to
$\hat m+\delta m$ with
$\delta m=-3g\hat m/2\pi$ in the second order term of (\ref{Sdem}). Note
that  although it makes no difference to enforce the gap equation to order
$g^3$ in the pressure (because of the cancellation discussed above),
there is no such cancellation in the entropy.

\subsection{Approximately self-consistent solutions}
\label{secscnum}

As we have seen, the 2-loop $\Phi$-derivable approximation
provides an expression for the entropy ${\cal S}$ as a functional
of the self-energy $\Pi$ which has a simple quasiparticle
interpretation and is manifestly ultraviolet finite for any
(finite) $\Pi$. These attractive
features of the formula (\ref{Ssc}) are independent of the specific
form of the self-energy, and will be shown to hold in QCD as well.
Of course, within this approximation, the self-energy
is uniquely specified: by the stationarity principle, this is
given by the self-consistent solution to the one-loop gap equation.
In the scalar $\phi^4$-model, it was easy to give the exact
solution to this equation.
In QCD, however, it will turn out that a fully self-consistent
solution is both prohibitively difficult (because of the
non-locality of the gap equation), and not really desirable
(because gauge symmetry implies relations between the
renormalization of the propagators and that of the vertices, and the
present approximation deals only with propagator renormalization).
This leads us to consider {\it approximately self-consistent}
resummations, which are obtained in two steps: (a) An approximation
is constructed for the solution
$\Pi$ to the gap equation, and (b) the entropy (\ref{Ssc}) is
evaluated {\it exactly} (i.e., numerically) with this approximate
self-energy. While step (b) above is unambiguous and inherently
non-perturbative, step (a), on the other hand, will be constrained
primarily by the requirement of preserving the maximum possible
perturbative accuracy, of order
$g^3$. In addition to that, we shall add the qualitative requirement
that the approximation for $\Pi$, and the ensuing one for ${\cal S}$,
are well defined and physically meaningful for all the values of $g$
of interest, and not only for small $g$---that is, for all the values
of
$g$ where the fully self-consistent calculation makes sense a priori.
Finally, in the case of QCD, relaxing the requirement of complete
self-consistency allows us to construct gauge invariant
approximations. 

\begin{figure}\label{phiqcd}
\includegraphics[width=.75\textwidth]{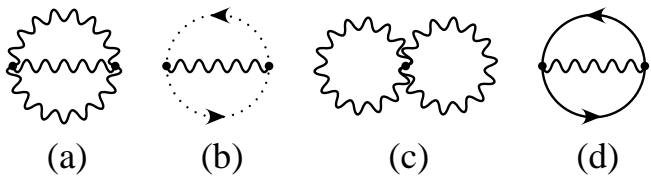}
\caption{QCD skeletons contributing to $\Phi$ at 2-loop order.
Wiggly, plain and dotted lines refer respectively to gluons, quarks
and ghosts}
\end{figure}

We shall now, in the rest of this lecture, outline the main steps that
are involved in the implementation of these approximations in the
case of QCD. Details can be found in the original publications
\cite{Blaizot:1999ip,Blaizot:1999ap,Blaizot:2001fc}.

At 2-loop order, the relevant skeletons are displayed in
Fig.~\ref{phiqcd}.  
By itself, the corresponding self-consistent truncation is not
a gauge invariant approximation. Our strategy then will be to 
use gauge-invariant
approximations to self-energies, in place of the self-consistent
ones. These self-energies are then used to compute the entropy 
without further approximations. In complete analogy with the
example of the scalar case that we have discussed in the previous
section, these approximations are such that, when expanded in powers
of the coupling the entropy is identical to that given by perturbation
theory up to and including order $g^3$.  

\begin{figure}\label{dPit}
\includegraphics[width=.75\textwidth]{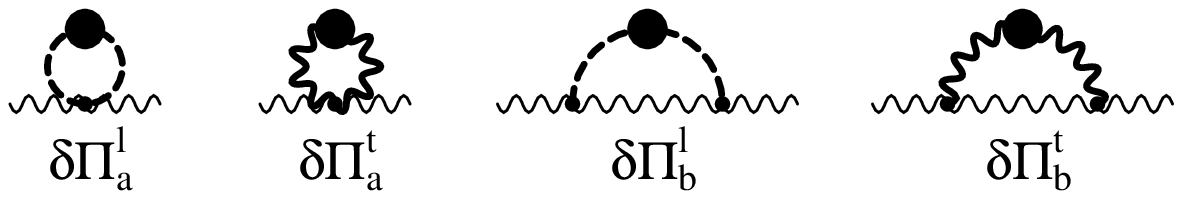}
\includegraphics[width=.5\textwidth]{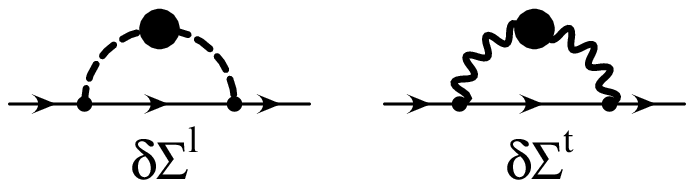}
\caption{Next to leading order contribution to $\delta \Pi_T$ (top) and to
$\delta \Sigma$ (bottom) at hard momentum. Thick dashed and wiggly lines
with a blob represent HTL-resummed longitudinal and transverse propagators,
respectively }
\end{figure}

The approximate self-energies that we  use are the hard thermal
loops\index{hard thermal loop} discussed above. Namely, for soft momenta
$\omega,\,p\sim gT$, we take  $\Pi_{soft}\,\approx\,\Pi_{HTL}$ and
$\Sigma_{soft}\,\approx\,\Sigma_{HTL}$, for gluons and quarks respectively. We
shall also need  an approximation valid for $\omega,\,p\sim T$:
$\Pi_{hard}(\omega^2\sim p^2)$ and similarly for $\Sigma$. It turns out that
this is accurately given by the hard thermal loop, even though the momenta are
not soft \cite{Kraemmer94}. All these approximations
 are gauge invariant. The corresponding diagrams are displayed in
Figs.~\ref{dPit}.

We can then proceed exactly as in the scalar case. As a first approximation 
one may simply use the hard thermal loops $\Pi=\Pi_{HTL}$ and $\Sigma\sim
\Sigma_{HTL}$ at
  all
momenta;  we refer to the corresponding entropy as ${\cal S}={\cal S}_{HTL}$.
The perturbative content of this approximation is schematically  
${\cal O}(g^2)\,+\,,\frac{1}{4}\,{\cal O}(g^3)$; that is, the approximation
fully accounts or the order $g^2$, but reproduce only a quarter of the $g^3$
order, exactly as in the scalar case. In 
 the next-to-leading approximation, we correct the hard degrees of
freedom by their interaction with the soft modes. That is, we continue to use
the hard thermal loops at small momenta, but use at hard momenta the
corrections corresponding to the diagrams displayed in Fig.~\ref{dPit}. The
resulting approximation to the entropy,  
${\cal S}={\cal S}_{NLA}$
accounts then fully for the orders 
$g^2$ and $g^3$. But of course these expressions  are not limited to values of
the coupling as small as required for the validity of perturbation theory.

\subsection{Some results for QCD}

\begin{figure}[h]\label{fig:pl1a}
\includegraphics[bb = 70 180 540 540,width=7.5truecm]{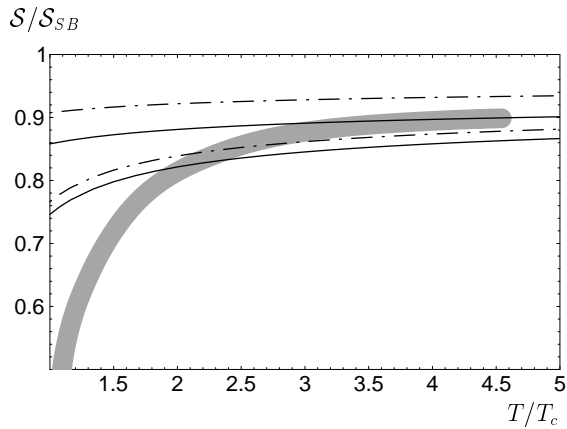}
\caption{The entropy of pure SU(3)
  gauge theory normalized to the ideal gas entropy
${\cal S}_0$.
  Full lines: ${\cal S}_{HTL}$.
Dashed-dotted lines: ${\cal S}_{NLA}$.
  2-loop $\beta$-function $\rightarrow$ the running
coupling constant $\alpha_s(\bar\mu)$.
The $\overline{{\rm MS}}$ renormalization scale:
  $\bar\mu=\pi T\cdots4\pi T$.
The dark grey band: lattice result by Boyd et al (1996) }
\end{figure}

As an illustration of the quality of the results that are obtained within that
scheme, we show in Fig.~\ref{fig:pl1a}  the entropy of pure SU(3)
gauge theory.  The bands delimiting the various lines in this figure
correspond to varying the $\overline{\rm MS}$ renormalization scale
$\bar\mu$, which defines the renormalized coupling constant $g(\bar\mu)$, from
$\bar\mu=\pi T$ to $4\pi T$. 
 One sees that in contrast to ordinary perturbation
theory, going from one level of approximation to the next one is indeed a small
correction. In particular the effects of the soft modes is here a
small contribution. This is o be contrasted with perturbation theory
where the  order
$g^3$ contribution is large for moderate values of the coupling. The
comparison with the lattice data \cite{Boyd} is quite good down to
$T\simge 2.5T_c$. 

The quality of the agreement between the self-consistent approximation and
the lattice data supports the quasiparticle picture of the quark-gluon plasma:
the dominant effect of the interactions at high temperature seems to be to
change the bare quarks and gluons into massive quasiparticles, with small
residual interactions between the quasiparticles. It should be emphasized that, 
in contrast to 
the approximations based on dimensional reduction,
the method  makes full use of the spectral information on the
quasiparticles contained in particular in the hard thermal loops. 

The approach is easily extended to finite chemical potential, and the
calculation of the baryonic density can be done using approximations similar to
those we used for the entropy. Furthermore, from the knowledge of $N(\mu,T)$ and
$S(\mu,T)$ one can reconstruct $P(\mu,T)$. Use lattice data to fix the
integration constant (e.g. $P(\mu=0,T)$). Such investigations are underway.

\section*{Acknowledgements}
I would like to thank the organizers of the school ``Dense Matter''
for what has been a very enjoyable meeting. I also 
express my gratitude to  Edmond Iancu and Anton Rebhan: much of the material
presented in these lectures is drawn from work done in collaboration.

%


\begin{thebibliography}{8.}
\addcontentsline{toc}{section}{References}



\bibitem{Blaizot:2001nr}
J.~Blaizot and E.~Iancu,
``The quark-gluon plasma: Collective dynamics and hard thermal loops,''
to appear in Phys. Rept, hep-ph/0101103.

\bibitem{Blaizot1997}
J.P. Blaizot,
{\it ``QCD at finite temperature''}, in ``Probing the Standard Model
of Particle Interactions'',
Les Houches, Session LXVIII, 1997, ed. by R. Gupta et al.
(Elsevier, Amsterdam, 1999).

%
\bibitem{Blaizot:1996ns}
J.~P.~Blaizot,
``The quark-gluon plasma and nuclear collisions at high energy,''
{\it Lecture given at Les Houches Summer School on Theoretical Physics, Session
66: Trends in Nuclear Physics, 100 Years Later, Les Houches, France, 
30 Jul - 30
Aug 1996}.


\bibitem{BIO95}
J.-P.~Blaizot, E.~Iancu and J.-Y.~Ollitrault, in {\it Quark Gluon Plasma 2},
  R.C.~Hwa ed.,
World Scientific, Singapore (1995), p. 135.


\bibitem{Puri94}
J.-P.~Blaizot, in {\it Lecture Notes on the Workshop: Nuclear Equation of
State}, Puri, India Jan. 1994, A.~Ansari and L.~Satpathy eds. (World 
Scientific,
Singapore, 1996).

\bibitem{Coree91}
J-P.~Blaizot, in Proceeding of Fourth Summer School and Symposium on
Nuclear Physics, {\it High Density and High Temperature Physics}, July
1-6, Namwon, Korea, Journ. Korean Phys. Soc., vol. 25.




\bibitem{KB62} L.~Kadanov and G.~Baym,
{\em Quantum Statistical mechanics}, (Benjamin/Cummings, London,1962).

\bibitem{Abrikosov63}
A.A.~Abrikosov, L.P.~Gorkov and I.E.~Dzyaloshinskii, {\em Methods of 
Quantum Field Theory in
Statistical Physics}, (Dover, New-York, 1963).

\bibitem{Fetter71}
A.~Fetter and J.D.~Walecka, {\em Quantum Theory of Many Particle 
Systems}, (McGraw Hill, New-York,
1971).

\bibitem{BR86} J.P.~Blaizot and G.~Ripka, {\em Quantum Theory
of Finite Systems}, (MIT Press, Cambridge, 1986).

\bibitem{Kapusta}
J.I.~Kapusta, {\it Finite temperature field theory}, Cambridge Monographs in
Mathematical Physics, (Cambridge University Press, 1989).


\bibitem{LeBellac96}
M.~Le Bellac, {\it Thermal field theory}, Cambridge Monographs in
Mathematical Physics, (Cambridge University Press, 1996).

\bibitem{Blaizot:1999ip}
J.~P.~Blaizot, E.~Iancu and A.~Rebhan,
Phys.\ Rev.\ Lett.\  {\bf 83} (1999) 2906
[hep-ph/9906340].


\bibitem{Blaizot:1999ap}
J.~P.~Blaizot, E.~Iancu and A.~Rebhan,
quark-gluon  plasma,''
Phys.\ Lett.\ B {\bf 470} (1999) 181
[hep-ph/9910309].


\bibitem{Blaizot:2001fc}
J.~P.~Blaizot, E.~Iancu and A.~Rebhan,
of  the quark-gluon plasma. I: Entropy and density,''
Phys.\ Rev.\ D {\bf 63} (2001) 065003
[hep-ph/0005003].



\bibitem{bagmodel}
K.~Johnson, C.B.~Thorn, A.~Chodos, R.L.~Jaffe  and V.F.~Weisskopf,
\PRD{9}{1974}{3471}; K.~Johnson, A.~Chodos, R.L.~Jaffe and C.B.~Thorn,
\PRD{10}{1974}{2599}.







\bibitem{Gins80}
P. Ginsparg, \NPB{170}{1980}{388}.

\bibitem{Appel81}
T. Appelquist and R.D. Pisarski, \PRD{23}{1981}{2305}.


\bibitem{Nadkarni83}
S. Nadkarni, \PRD{27}{1983}{917}; \PRD{38}{1988}{3287}.

\bibitem{Landsman89}
N.P. Landsman, \NPB{322}{1989}{498}.

\bibitem{Braaten94}
E. Braaten, \PRL{74}{1995}{2164}.

\bibitem{Kajantie94}
K. Kajantie, K.~Rummukainen and M.E.~Shaposhnikov,
\NPB{407}{1993}{356};
K. Farakos, K. Kajantie, K.~Rummukainen and M.E.~Shaposhnikov,
\NPB{425}{1994}{67}; {\it ibidem} {\bf 442} (1995) 317.

\bibitem{Linde79}
A.D. Linde, Rep. Progr. Phys. {\bf 42} (1979) 389.

\bibitem{Linde80}
  A.D.~Linde, \PLB{96}{1980}{289}.


\bibitem{qed}
J.P. Blaizot and E. Iancu, \NPB{390}{1993}{589}.

\bibitem{qcd}
J.P. Blaizot and E. Iancu, \PRL{70}{1993}{3376};
\NPB{417}{1994}{608}.


\bibitem{Klimov81}
V.V.~Klimov, \SovJNuclPhys{33}{1981}{934}; \SovPhysJETP{55}{1982}{199}.

\bibitem{Weldon82a}
  H.A.~Weldon,\PRD{26}{1982}{1394}.

\bibitem{Weldon82b}
  H.A.~Weldon,\PRD{26}{1982}{2789}.

\bibitem{Pisarski89}
R. D.~Pisarski, \PRL{63}{1989}{1129}; E.~Braaten and R. D.~Pisarski,
\NPB{337}{1990}{569}; \NPB{339}{1990}{310};
  \PRL{64}{1990}{1338}; \PRD{42}{1990}{2156}.

\bibitem{FT90}
J.~Frenkel and J.C.~Taylor,\NPB{334}{1990}{199};
J.C. Taylor and S.M.H. Wong, {\it ibid.} {\bf B346} (1990) 115.

\bibitem{Nair91}
R.~Efraty and V.P.~Nair, \PRL{68}{1992}{2891};
R.~Jackiw and V.P.~Nair,  \PRD{48}{1993}{4991}.



\bibitem{PhysKin}
E.M.~Lifshitz and L.P.~Pitaevskii, {\it Physical Kinetics} (Pergamon Press,
Oxford, 1981).

\bibitem{Silin60}
V.P.~Silin, \SovPhysJETP{11}{1960}{1136}.


\bibitem{Wigner1}
N.L. Balazs and B.K. Jennings, \PhysRept{104}{1984}{347}.

\bibitem{Wigner2}
M. Hillery, R.F. O'Connell, M.O. Scully, and E.P. Wigner,
\PhysRept{106}{1984}{121}.



\bibitem{BP91}
G. Baym and C.J. Pethick, {\it Landau Fermi-liquid theory:
concepts and applications}, (J. Wiley, N.Y., 1991).

\bibitem{Wong}
S.K. Wong, Nuovo Cimento {\bf 65A} (1970) 689.


\bibitem{Heinz83}
U.~Heinz, \PRL{51}{1983}{351}; J.~Winter, J.Phys. (Paris) Suppl.{\bf 45}
(1984) C4-53;  U.~Heinz,  \AnnPhys{161}{1985}{48}.

\bibitem{EHPRept}
H.-Th.~Elze and U.~Heinz, \PhysRept{183}{1989}{81}.

\bibitem{Kelly94}
P.F. Kelly, Q. Liu, C. Lucchesi and C. Manuel, \PRL{72}{1994}{3461};
\PRD{50}{1994}{4209}.

\bibitem{Gyulassy93}
A. V. Selikhov and M. Gyulassy, \PLB{316}{1993}{373};
\PRC{49}{1994}{1726}.

\bibitem{Markov95}
Yu.A. Markov, M.A. Markova, Theor. Math. Phys. {\bf 103} (1995) 444.


\bibitem{Kraemmer94}
U. Kraemmer, A. Rebhan and H. Schulz, \AnnPhys{238}{1995}{286}.


\bibitem{prl2}
J.P. Blaizot and E. Iancu,   \PRL{72}{1994}{3317}.

\bibitem{planew}
J.P. Blaizot and E. Iancu,
\PLB{326}{1994}{138}.


\bibitem{Baym90}
G.~Baym, H.~Monien, C.J.~Pethick, and D.G.~Ravenhall,
\PRL{64}{1990}{1867}.


\bibitem{Arnold94}
P.~Arnold and C.~Zhai, \PRD{50}{1994}{7603};\PRD{51}{1995}{1906}; C.~Zhai and
B.~Kastening, \PRD{52}{1995}{7232}. See also, E.~Braaten, \PRL{74}{1995}{2164};
E.~Braaten and A.~Nieto, \PRD{51}{1995}{6990}.

\bibitem{Braaten:1995na}
E.~Braaten,
Phys.\ Rev.\ Lett.\  {\bf 74} (1995) 2164
[hep-ph/9409434].

\bibitem{Braaten:1996ju}
E.~Braaten and A.~Nieto,
Phys.\ Rev.\ Lett.\  {\bf 76} (1996) 1417
[hep-ph/9508406].

\bibitem{Braaten:1996jr}
E.~Braaten and A.~Nieto,
Phys.\ Rev.\ D {\bf 53} (1996) 3421
[hep-ph/9510408].


\bibitem{Kajantie:2001iz}
K.~Kajantie, M.~Laine, K.~Rummukainen and Y.~Schroder,
Phys.\ Rev.\ Lett.\  {\bf 86} (2001) 10
[hep-ph/0007109].



\bibitem{Hatsuda:1997wf}
T.~Hatsuda,
Phys.\ Rev.\ D {\bf 56} (1997) 8111
[hep-ph/9708257].

\bibitem{Chiku:1998kd}
S.~Chiku and T.~Hatsuda,
Phys.\ Rev.\ D {\bf 58} (1998) 076001
[hep-ph/9803226].


\bibitem{Parwani:2001rr}
R.~R.~Parwani,
Yang-Mills  theory,''
Phys.\ Rev.\ D {\bf 63} (2001) 054014
[hep-ph/0010234].

\bibitem{Parwani:2001am}
R.~R.~Parwani,
Phys.\ Rev.\ D {\bf 64} (2001) 025002
[hep-ph/0010294].





\bibitem{Peshier}A. Peshier, B. K\"ampfer, O. P. Pavlenko, and G. Soff,
         Phys. Rev. D {\bf 54}, 2399 (1996); A. Peshier, {\tt hep-ph/9809379}.

\bibitem{LH} P. L\'evai and U. Heinz, Phys. Rev. C {\bf 57}, 1879 (1998)
         and references therein.
\bibitem{Karsch97}
F.~Karsch, A.~Patkos and P.~Petreczky,\PLB{401}{1997}{69}.

\bibitem{ABS} J. O. Andersen, E. Braaten, and M. Strickland,
         Phys. Rev. Lett. {\bf 83}, 2139 (1999), 
         Phys. Rev. D {\bf 61}, 014017, 074016 (2000).







\bibitem{LW} J. M. Luttinger and J. C. Ward, Phys. Rev. {\bf 118}, 1417
         (1960); C. De Dominicis and P.C. Martin, J. Math. Phys. {\bf 5},
         14, 31 (1964).
\bibitem{Baym} G. Baym, Phys. Rev. {\bf 127}, 1391 (1962).
\bibitem{Riedel} E. Riedel, Z. Phys. {\bf 210}, 403 (1968).
\bibitem{VB} B. Vanderheyden and G. Baym, J. Stat. Phys. {\bf 93}, 843 (1998).
\bibitem{CP2}G. M. Carneiro and C. J. Pethick,
         Phys. Rev. B {\bf 11}, 1106 (1975).


\bibitem{Drummond:1998cw}
I.~T.~Drummond, R.~R.~Horgan, P.~V.~Landshoff and A.~Rebhan,
Nucl.\ Phys.\ B {\bf 524} (1998) 579
[hep-ph/9708426].


\bibitem{Boyd} G. Boyd {\it et al.},
        Nucl. Phys. {\bf B469}, 419 (1996).        










\end{thebibliography}
\end{document}